\documentclass[twocolumn, superscriptaddress, aps, pra, floatfix]{revtex4-2}

\usepackage{color}
\usepackage{graphicx}
\usepackage{float}
\usepackage{dcolumn}
\usepackage{braket}
\usepackage{physics}
\usepackage{amsmath,amssymb}
\newcommand{\figuref}[1]{\mbox{Figure~\ref{#1}}}
\usepackage[breaklinks=true,colorlinks,citecolor=blue,linkcolor=blue,urlcolor=blue]{hyperref}
\usepackage[capitalise]{cleveref}
\usepackage{makeidx}

\usepackage[dvipsnames]{xcolor}

\newcommand{\figref}[1]{\mbox{Fig.~\ref{#1}}}
\newcommand{\figpanel}[2]{Fig.~\hyperref[#1]{\ref*{#1}(#2)}}
\newcommand{\figurepanel}[2]{Figure~\hyperref[#1]{\ref*{#1}(#2)}}
\newcommand{\figpanels}[3]{Figs.~\hyperref[#1]{\ref*{#1}(#2)-(#3)}}

\newcommand{\be}{\begin{equation}}
\newcommand{\ee}{\end{equation}}
\newcommand{\bea}{\begin{eqnarray}}
\newcommand{\eea}{\end{eqnarray}}

\renewcommand{\eqref}[1]{\mbox{Eq.~(\ref{#1})}}

\newcommand{\expec}[1]{\left\langle{#1}\right\rangle}

\begin{document}

\author{Gabriele Orlando}
\affiliation{Dipartimento di Scienze Matematiche e Informatiche, Scienze Fisiche e  Scienze della Terra, Universit\`{a} di Messina, I-98166 Messina, Italy}

\author{Daniele Lamberto}
\email{daniele.lambe@gmail.com}
\affiliation{Dipartimento di Scienze Matematiche e Informatiche, Scienze Fisiche e  Scienze della Terra, Universit\`{a} di Messina, I-98166 Messina, Italy}
\affiliation{Quantum Information Physics Theory Research Team, RIKEN Center for Quantum Computing, Wakoshi, Saitama, 351-0198, Japan}

\author{Franco Nori}
\affiliation{Quantum Information Physics Theory Research Team, RIKEN Center for Quantum Computing, Wakoshi, Saitama, 351-0198, Japan}
\affiliation{Physics Department, The University of Michigan, Ann Arbor, Michigan 48109-1040, USA}

\author{Salvatore Savasta}
\affiliation{Dipartimento di Scienze Matematiche e Informatiche, Scienze Fisiche e  Scienze della Terra,	Universit\`{a} di Messina, I-98166 Messina, Italy}

\title{Quantum Vacuum Radiation Near a Critical Point}

\begin{abstract}
Equilibrium quantum phase transitions profoundly reshape the ground state of light-matter systems; yet, the resulting quantum correlations, such as squeezing and entanglement, remain experimentally inaccessible since they involve virtual ground state excitations. We investigate how a nonadiabatic modulation of a Hamiltonian parameter can convert these virtual excitations into real photons, enabling quantum vacuum radiation. We show that proximity to the critical point strongly enhances the emitted photon flux and the non-classical nature of the emitted radiation, even when thermal fluctuations are expected to dominate. In addition, higher-order processes become relevant even for small modulation amplitudes, and we develop a framework that systematically incorporates them. Our results reveal that criticality can act as an efficient amplifier of vacuum fluctuations, offering new routes to probe and exploit quantum critical ground states.
\end{abstract}

\maketitle

\section{Introduction}

Quantum correlations, such as squeezing and entanglement, constitute key signatures of non-classical behavior and represent valuable resources for quantum technologies \cite{Horodecki2009entanglementRMP, Weedbrook2012gaussianRMP, Braunstein2005QuantuminfoRMP, Schnabel2017squeezinglight, Kitagawa1993squeezingspin, MaNori2011squeezingspin, NielsenChuang_book}. However, their direct detection in many-body systems is often challenging \cite{Amico2008entanglementmanybodyRMP, PlenioVirmani2005entanglementmeasures, Sorensen2001entanglementBoseEinstein, Eisert2010arealawsRMP, Sachdev2008magnetismcriticality}. In several platforms, the most interesting correlations are encoded in the ground state and involve virtual excitations that cannot be directly observed through standard spectroscopic probes \cite{Ciuti2005vacuumintersubbandpolariton, DeLiberato2007vacuumradiation}. 
This raises the central question of how quantum correlations can be experimentally accessed, since they are often hidden in the ground state and of weak magnitude. 

A promising route emerges in the vicinity of quantum phase transitions (QPTs). QPTs correspond to qualitative changes in the ground state of a quantum system driven by quantum fluctuations at zero temperature and controlled by a non-thermal parameter \cite{Sachdev_book, Carr2010book, Sondhi1997qptsRMP, Vojta2003qpts}. Near the critical point, the system develops long-range correlations and enhanced fluctuations, which can lead to drastic enhancements and modifications of its quantum properties \cite{Osterloh2002entanglementqpt, Osborne2002entanglementqpt, Vidal2003entanglementcritical, Vidal2004entanglementLMGqpt}. In particular, quantum correlations can display scaling behavior or even divergences when approaching criticality. As a consequence, critical points naturally act as amplifiers of non-classical features \cite{Hertz1976quantumcritical, Nataf2012entanglementspt}.

This behavior has been extensively discussed in a variety of models exhibiting second-order QPTs. Illustrative examples include the transverse Ising and Heisenberg models \cite{Pfeuty1970Isingqpt, Botet1982Isingqpt, Dyson1976Heisenbergqpt, Roche2021photoncondensation, Schaffer2012HeisenbergKitaevqpt,  Dutta_book}, the Lipkin-Meshkov-Glick model \cite{Lipkin1965LMGqpt, Vidal2004entanglementLMGqpt, Tsomokos_2008}, the Bose-Hubbard model \cite{Fisher1989BoseHubbardqpt, Greiner2002BoseHubbardqpt}, and the Dicke, Rabi and renormalized Hopfield models \cite{Dicke1954spontaneouscoherence, HeppLieb1973Dickeqpt, Wang1973Dickeqpt, Hioe1973generalizedDickeqpt, Brandes2003DickeqptPRE, Brandes2003DickeqptPRL, Hwang2015Rabiqpt, Roche2025, chen2024sudden, Lamberto2025manydipoleqpt}. In particular, it has been shown that quantities such as entanglement entropy can diverge at criticality \cite{Lambert2004entanglementDicke}, while quantum squeezing (corresponding to a redistribution of quantum uncertainty between complementary observables) can become increasingly pronounced when the system approaches the critical point \cite{Brandes2003DickeqptPRE, Hayashida2023squeezingDicke}. In this perspective, quantum correlations are not only signatures of critical phenomena but also potential resources for criticality-enhanced quantum technologies and sensing applications \cite{Chu2021criticalquantumsensing, Ilias2022criticalquantumsensing, Hotter2024criticalquantummetrology, Hotter2025metrology, Degen2017quantumsensingRMP,Liu2016Metrology,cirio2017amplified}.

Despite these appealing properties, the quantum correlations present in the ground state near a critical point are typically not directly accessible experimentally, as they involve virtual excitations that remain bound to the system. When such systems are probed using weak coherent drives, the response is governed by the polaritonic normal modes, which behave as effectively non-interacting harmonic oscillators. As a result, the system behaves as a linear optical medium, and standard spectroscopic measurements reveal only limited signatures of the approaching phase transition, such as the softening of the lowest-energy polariton mode \cite{Lamberto2025manydipoleqpt,Kono2025observationspt, Hotter2025quantum}.

\begin{figure}[t]
    \centering
    \includegraphics[width=\columnwidth]{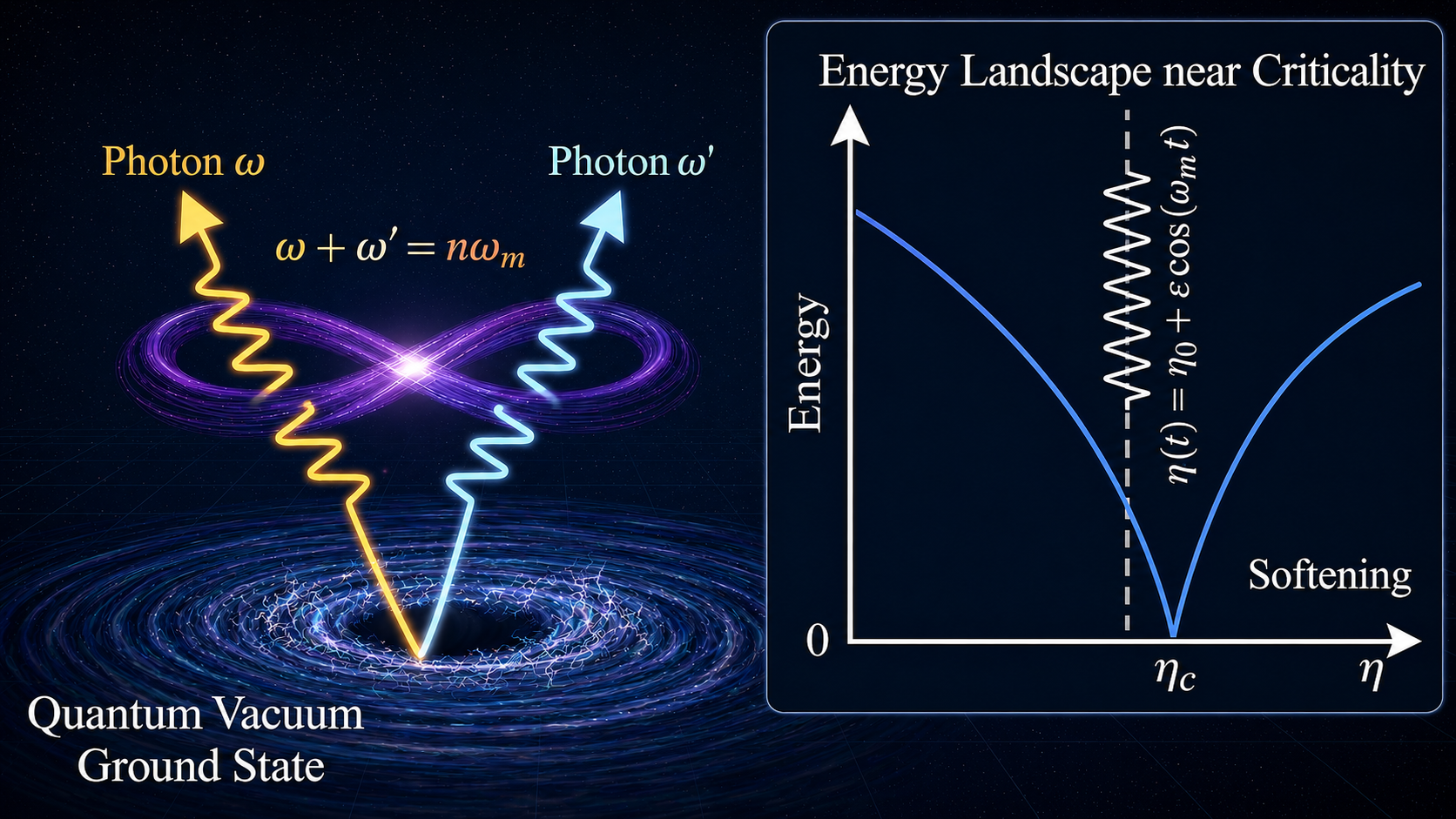}
    \caption{\textbf{Quantum vacuum emission near the critical point.}
    Schematic representation of the process under investigation. A nonadiabatic modulation of a Hamiltonian parameter induces the emission of a pair of photons with frequencies $\omega$ and $\omega'$ from the vacuum ground state, satisfying the resonance condition $\omega+\omega'=n\omega_m$ (see \cref{sec:photon_flux_density}). The intertwined purple ribbon schematically denotes the entanglement between the emitted photons, discussed in \cref{sec:entanglement}. The inset on the right represents a schematic dispersion relation in the vicinity of the critical coupling $\eta_c$, showing the characteristic mode softening. The vertical wavy line represents the periodic modulation of the coupling parameter, $\eta(t)=\eta_0+\epsilon\cos(\omega_m t)$ (see \cref{Sec:model}).}
   
    \label{fig:model}
\end{figure}

To overcome this limitation, we propose a nonadiabatic time modulation of one of the Hamiltonian parameters controlling the phase transition. This modulation converts virtual excitations and the quantum correlations present in the ground state into real, detectable ones \cite{Ciuti2005vacuumintersubbandpolariton,DeLiberato2007vacuumradiation}. In this way, correlations that are otherwise hidden in the ground state can be transferred into observable radiation. A schematic representation of the process is presented in \figref{fig:model}.

The possibility of generating particles from the quantum vacuum is a direct consequence of the Heisenberg uncertainty principle, which allows vacuum fluctuations to manifest as real excitations under suitable conditions. Several physical mechanisms illustrate this phenomenon. Parametric amplification can convert vacuum fluctuations into real photons and is widely used in quantum optics and superconducting circuits \cite{Clerk2010noiseamplificationRMP}. The dynamical Casimir effect shows that rapidly varying boundary conditions, such as oscillating mirrors, can produce photons from the vacuum \cite{Moore1970dce,  Johansson2009dcecircuitPRL, Johansson2010dcecircuitPRA,
Wilson2011observationdce, dodonov2020fifty, macri2018nonperturbative}. Related concepts also appear in high-energy and gravitational physics \cite{Nation2012colloquium}, including Hawking radiation \cite{Hawking1974blackhole}, the Unruh effect \cite{Unruh1976blackhole}, and the Schwinger mechanism \cite{Schwinger1951vacuumpolarization}, where strong fields or accelerated frames lead to particle creation from vacuum fluctuations. 

In light-matter systems operating in the ultrastrong coupling regime \cite{FriskKockum2019ultrastrongNRP, FornDiaz2019ultrastrongRMP,garcia-vidal2021manipulating}, the ground state itself contains virtual quantum-correlated excitations.  Nonetheless, these excitations remain bound to the system and cannot escape as radiation \cite{Savasta1996inputoutput,DeLiberato2017virtual}.
However, it has been shown that such virtual excitations can be released if the system parameters are varied sufficiently rapidly in time \cite{DeLiberato2007vacuumradiation,deliberato2009extracavity, Garziano2013switching, Stassi2013spontaneous,garziano2014vacuum,SanchezMunoz2018,Giannelli2024detecting}. In particular, a nonadiabatic modulation of the light-matter coupling rate can generate radiation from the quantum vacuum in close analogy with the dynamical Casimir effect \cite{DeLiberato2007vacuumradiation,Johansson2009dcecircuitPRL,Johansson2010dcecircuitPRA,Wilson2011observationdce,carusotto2008optical}.

In this work we investigate how quantum vacuum radiation is affected by the proximity to a quantum critical point, at both zero and finite temperature. In particular, we show that the vicinity of the critical point strongly enhances the emitted photon flux and qualitatively modifies the emission processes. Moreover, we explicitly illustrate how measurable non-classical indicators, such as entanglement and squeezing, are exponentially enhanced by criticality. We also demonstrate that, when approaching the critical point, higher-order resonance processes become increasingly important, and the standard linear description is no longer sufficient. 

To address this regime, we develop a theoretical framework based on the Quantum Langevin Equations (QLEs), without relying on the Born-Markov approximation, thereby allowing us to systematically account for the contributions of higher-order processes. Indeed, in the presence of criticality this approximation breaks down, as the system-bath interaction becomes progressively stronger (the ratio between the system losses and its characteristic frequencies inevitably grows), thus preventing the use of standard techniques such as the master-equation approach \cite{Lamberto2026superradiantquantum}. Crucially, we identify instability regions in the proximity of the critical point where the mean field solutions exhibit an exponential growth in the time domain.
As a concrete example, we consider a system which can be regarded as the dispersive limit of the well-known Dicke model, which exhibits a superradiant quantum phase transition (SPT). While the specific model serves as an illustrative platform, our focus is on the general role played by quantum criticality in enhancing vacuum emission and revealing ground-state correlations.

The paper is organized as follows. In \cref{Sec:model}, we introduce the system under consideration and present the theoretical framework based on QLEs, together with the notation adopted throughout the manuscript. In \cref{sec:photon_flux_density} and \cref{sec:photon_flux}, we investigate vacuum emission through the analysis of the photon-flux density and the photon flux, respectively, at both zero and finite temperatures, illustrating the main features of the emission spectra with representative examples. In \cref{sec:squeezing}, we discuss the squeezing properties of the quantum vacuum radiation induced by the nonadiabatic time modulation, through the use of the squeezing spectrum and Wigner representation, and their experimental observability. Finally, in \cref{sec:entanglement}, we explore the nonclassical properties of the vacuum emission and the entanglement between pairs of modes, with particular emphasis on the behavior of the system as it approaches the critical point.

\section{Theoretical framework}\label{Sec:model}

In this work, we consider a widely employed, bosonic Hamiltonian, which displays a softening of its eigenmode (namely, the vanishing of the excitation energy related to the lowest eigenmode), signaling the presence of a QPT. The Hamiltonian that we will consider throughout this work is given by
\begin{equation} \label{eq:H_sys_dispersive}
    H_\mathrm{sys} = \hbar \omega_a a^\dagger a - \hbar \eta \left( a + a^\dagger \right)^2 \, ,
\end{equation}
where $\omega_a$ is the bosonic frequency and $\eta$ is the coupling strength which dictates the interaction between the bosonic field and some higher-energy levels.
In particular, the Hamiltonian in \cref{eq:H_sys_dispersive} has been extensively used since it can be interpreted as the dispersive limit of the Dicke model in the normal phase, whenever the transition frequency of the two-level systems $\omega_b$ is much greater than the bosonic frequency, i.e., $\omega_b \gg \omega_a$ \cite{Zueco2009dispersive}.
As can be readily verified, this Hamiltonian exhibits a QPT at the critical coupling \(\eta_c=\omega_a/4\). This value matches the critical point of the Dicke model in the thermodynamic limit, \(g_c=\sqrt{\omega_a\omega_b}/2\), since \(\eta=g^2/\omega_b\) in the dispersive limit. This relation enables direct experimental access and it is relevant for the subsequent results, as nonadiabatic periodic modulations of $\omega_b$ can be implemented through a time-dependent applied magnetic field; while, on the contrary, a direct nonadiabatic modulation of the coupling strength \(g\) is usually more difficult to realize.

We now develop a comprehensive theoretical framework for the corresponding open system, able to give reliable predictions even near the critical point, where the usual approximations (e.g., Born-Markov) based on the smallness of the ratio between the losses and the eigenfrequencies of the system fail.
To this end, we model the external environment as an infinite collection of independent harmonic oscillators and, thus, the total Hamiltonian (system plus external bath) can be written as \cite{GardinerZoller_book}
\begin{equation} \label{eq:H_tot}
    H_\mathrm{tot} = H_\mathrm{sys} + \frac{1}{2}\displaystyle \sum_n \left[ p_{n}^2 + k_{n}(q_{n} - X_a)^2 \right]
\end{equation}
where $q_{n}$ and $p_{n}$ are the coordinate and momentum associated to the $n$-th mode of the bath, respectively, and $X_a = \sqrt{\hbar/2\omega_a}(a^\dagger + a)$ is the system coordinate.
Building on the standard quantum Langevin and input-output formalisms~\cite{Scully_book,WallsMilburn_book_input-output,GardinerZoller_book,collett1984squeezing,gardiner1985input}, and following the approach of Ref.~\cite{Lamberto2026superradiantquantum}, we derive from \eqref{eq:H_tot} the quantum Langevin equations (QLEs) and the corresponding input-output relations.
In particular, the QLEs for the bosonic operators read (see \cref{app:derivation_langevin_input})
\begin{align} \label{eq:QLE_time}
    \dot{a}(t) = & -i\omega_a a(t) + 2i \eta(t) \left[ a(t) + a^\dagger (t) \right] +\frac{i}{\sqrt{2\hbar\omega_a}} \xi(t) \nonumber \\
    & -\frac{i}{2\omega_a}\int_{t_0}^t \! f(t-t')[\dot{a}^\dagger(t')+\dot{a}(t')] dt' \, , 
\end{align} 
where $\xi(t)$ is the Hermitian bath noise operator, while $f(t)$ plays the role of a memory function for the bath. The QLE for  $a^\dagger(t)$ follows straightforwardly by taking the Hermitian conjugate of \eqref{eq:QLE_time}.

We now apply the Fourier transform to both sides, and consider for the input fields the initial time $t_0$ to be in the distant past, i.e., $t_0 \to - \infty$. Therefore, we obtain the following equation

\begin{align} \label{eq:QLE_frequency}
    -i \omega \tilde{a}(\omega) = & -i\omega_a \tilde{a}(\omega) + 2i \tilde{\eta}(\omega) * \left( \tilde{a}(\omega) + \tilde{a}^\dagger(\omega) \right)  \nonumber \\
    & + \frac{i}{\sqrt{2\hbar\omega_a}} \tilde{\xi}(\omega) + \frac{\gamma (\omega)}{2} \left( \tilde{a}^\dagger(\omega) - \tilde{a}(\omega) \right) \, , 
\end{align} 
where the symbol $*$ denotes the convolution operation, namely
$
\tilde{A}(\omega)*\tilde{B}(\omega)
=\int_{-\infty}^{+\infty}\! d\omega'\,\tilde{A}(\omega')\,\tilde{B}(\omega-\omega'),
$ and \(\gamma(\omega)\) is the decay rate function; further details are provided in \cref{app:derivation_langevin_input}. We emphasize that since the treatment adopted here goes beyond the Markov approximation, the damping rate \(\gamma(\omega)\) may in general acquire an explicit frequency dependence, whereas for the ohmic bath (considered throughout the paper) it reduces to a constant \cite{Lamberto2026superradiantquantum, Leggett1987dissipativeRMP}.

We now consider a periodic modulation of frequency $\omega_m$ of the coupling strength given by $\eta (t) = \eta_0 + \epsilon \cos \left( \omega_m t \right)$. Normally, when no critical point is present near the region of interest, it is sufficient to require the modulation amplitude $\epsilon$ to be sufficiently small (namely, $\epsilon \ll \eta_0$) to apply a linear theory and have reliable predictions. However, this is not the case  near a critical point, even if the modulation is relatively small, as we will demonstrate below.

As it can be noticed in \cref{eq:QLE_frequency}, the modulation term in the frequency domain is given by the convolution of the Fourier transforms $\tilde{\eta} (\omega)$ and $\tilde{a}(\omega)$. Given that a periodic time modulation of frequency $\omega_m$ has been assumed for the coupling strength,the resulting modulation of $\tilde{a}(\omega)$ is given explicitly by
\begin{equation} \label{eq:modulation_frequency_domain}
    \tilde{\eta}(\omega) * \tilde{a}(\omega) = \eta_0 \tilde{a}(\omega) + \frac{\epsilon}{2} \left[ \tilde{a}(\omega + \omega_m) + \tilde{a}(\omega - \omega_m) \right] \, ,
\end{equation}
with an analogous relation holding for $\tilde{a}^\dagger (\omega)$.
Therefore, the QLEs can be compactly written as
\begin{align} \label{eq:QLE_input_extended}
    - i \omega \tilde{\mathbf{v}} (\omega) =& - i \left(\mathbf{A} - \frac{i}{2} \mathbf{\Gamma}(\omega) \right) \tilde{\mathbf{v}}(\omega) \nonumber \\
    &+ i \mathbf{A}_\epsilon \left[ \tilde{\mathbf{v}} (\omega + \omega_m) + \tilde{\mathbf{v}} (\omega - \omega_m) \right] + \tilde{\mathbf{F}}_\mathrm{in} (\omega)\,,
\end{align}
where $\tilde{\mathbf{v}} (\omega) = \left( \tilde{a}(\omega), \tilde{a}^\dagger (\omega) \right)^T$ and $\tilde{\mathbf{F}}_{\mathrm{in}}(\omega)$ is the Langevin force vector of the input fields in the frequency domain, while $\mathbf{A}$ and $\mathbf{\Gamma}$ are the Bogoliubov and decay matrices of the non-modulated system, respectively, and $\mathbf{A}_\epsilon$ is the modulation matrix, which are given by
\begin{align} \label{eq:A_Gamma_matrices}
    \mathbf{A} &= \begin{pmatrix}
        \omega_a - 2 \eta_0 & - 2 \eta_0 \\
        2 \eta_0 & - \omega_a + 2 \eta_0 
    \end{pmatrix} \, , \\
    \mathbf{\Gamma} &= \begin{pmatrix}
        \gamma & - \gamma \\
        - \gamma & \gamma 
    \end{pmatrix} \qquad\qquad\quad\,\,\,\ , \\
    \mathbf{A}_\epsilon &= \begin{pmatrix}
        \epsilon & \epsilon \\
        -\epsilon & -\epsilon
    \end{pmatrix} \qquad\qquad\quad\quad\, .
\end{align}
By introducing the matrix $\mathbf{M}(\omega) = \mathbf{A} - i \mathbf{\Gamma} / 2 - \omega \mathbf{I}$, \cref{eq:QLE_input_extended} can be written as
\begin{equation} \label{eq:QLE_input}
    \mathbf{M}(\omega) \tilde{\mathbf{v}} (\omega) = \mathbf{A}_\epsilon \left[ \tilde{\mathbf{v}} (\omega + \omega_m) + \tilde{\mathbf{v}} (\omega - \omega_m) \right] - i \tilde{\mathbf{F}}_\mathrm{in} (\omega)\, .
\end{equation}

Following the approach outlined in Refs.~\cite{Yurke1984networktheory,GardinerZoller_book}, we introduce the creation and annihilation operators of the input fields as
\begin{equation} \label{eq:C_in}
     C_\mathrm{in}(t) = \int_0^\infty \sqrt{\frac{\hbar}{4 \pi \omega}} \left( \tilde{c}_\mathrm{in}(\omega) e^{-i \omega t} + \tilde{c}_\mathrm{in}^\dagger (\omega) e^{i \omega t} \right) d\omega \, ,
\end{equation}
where the bosonic operators satisfy the canonical commutation relations. 
The input Langevin force vector in the frequency domain is strictly linked to the bosonic operators of the input field as 
\begin{equation} \label{eq:F_in_main}
    \tilde{\mathbf{F}}_\mathrm{in} (\omega) = i \sqrt{ \frac{\abs{\omega} \gamma} {\omega_a}}\, \check{c}_\mathrm{in}(\omega) \begin{pmatrix}
        1 \\
        -1 
    \end{pmatrix} \, ,
\end{equation}
with the short notation
\begin{equation} \label{eq:c_check}
    \check{c}_\mathrm{in}(\omega) = \tilde{c}_\mathrm{in}(\omega) \theta(\omega) + \tilde{c}^\dagger_\mathrm{in}(-\omega) \theta(-\omega) \, .
\end{equation}
See \cref{app:derivation_langevin_input} for further details.

The QLEs for the output fields can be derived following a similar procedure, by considering $t_0 \to +\infty$, which leads to an analogous expression to \cref{eq:QLE_input_extended} where the sign of the decay matrix is reversed. Moreover, it can be easily shown that the output field is related to the input by the relation 
\begin{equation} \label{eq:output_vs_input}
    \tilde{\mathbf{F}}_\mathrm{out}(\omega) = \tilde{\mathbf{F}}_\mathrm{in}(\omega) - \mathbf{\Gamma}(\omega) \tilde{\mathbf{v}}(\omega) \, .
\end{equation}

By consistently solving \cref{eq:QLE_input} for the bosonic vector $ \tilde{\mathbf{v}}(\omega)$ and inserting the result into \cref{eq:output_vs_input}, we obtain the following expression for the output field at positive frequencies (see \cref{sec:derivation_output} for the derivations of this equation and of the corresponding coefficients)
\begin{align} \label{eq:out_in_definition}
    \tilde{c}_\mathrm{out}(\omega) &\!=\! k_0 (\omega) \check{c}_\mathrm{in}(\omega) +\! \sum_{n=1}^{+\infty} \left[ \vphantom{\tilde{c}_\mathrm{in}^\dagger} k_n (\omega) \check{c}_\mathrm{in}(\omega + n \omega_m) \right. \nonumber \\ 
    & \left. \qquad\qquad\qquad\quad\;\; \vphantom{\vphantom{\tilde{c}_\mathrm{in}^\dagger}} + k_{-n} (\omega) \check{c}_\mathrm{in}(\omega - n \omega_m) \right] .
\end{align}
While the first term in the summation is always associated to a annihilation operator, as it can be readily seen from \cref{eq:c_check} since $(\omega + n \omega_m) > 0$, the second term contains both creation and destruction operators depending on the sign of ($\omega - n \omega_m$) and is the term responsible for the quantum vacuum radiation.

To perform explicit calculations, we retain terms up to a sufficiently large \(N\) in \cref{eq:out_in_definition}, thereby including parametric processes involving up to the \(N\)-th harmonic of the modulation frequency. This truncation procedure is justified because the contributions associated with increasingly higher harmonics decay rapidly, becoming progressively negligible.






To make a comparison with the linear theory (output photon flux proportional to $\epsilon^2$), which is reliable for small modulation amplitudes far from the critical point, we firstly truncate the expansion of the output annihilation operator in \cref{eq:out_in_definition} up to the first harmonic $(N=1)$, which, for $\omega < \omega_m$, reads 
\begin{align} \label{eq:out_1_harm}
    \tilde{c}_{\rm out}(\omega) =& \left[ 1 - \tilde{k}(\omega) \right] \tilde{c}_{\rm in}(\omega) \nonumber \\
    & + \frac{2 \epsilon \omega_a \tilde{k}(\omega)}{D(\omega + \omega_m)} \sqrt{\frac{\omega + \omega_m}{\omega}} \tilde{c}_{\rm in}(\omega + \omega_m) \nonumber \\
    &  + \frac{2 \epsilon \omega_a \tilde{k}(\omega)}{D(\omega - \omega_m)} \sqrt{\frac{\omega_m - \omega}{\omega}} \tilde{c}_{\rm in}^\dagger(\omega_m - \omega) \, ,
\end{align}
where we defined the coefficients
\begin{align}
    \tilde{k}(\omega)\! &=\! \frac{2 i \gamma_a \omega}{D(\omega)}\! \left[ 1 \! - 4 \epsilon^2 \omega_a^2 \! \left( \! \frac{D(\omega + \omega_m) + D(\omega - \omega_m)}{D(\omega) D(\omega + \omega_m) D(\omega - \omega_m)} \! \right) \! \right]^{-1} \!\!\! , \\
    D(\omega) &= \det(M(\omega)) = \omega^2 + i \gamma_a \omega - \tilde{\omega}_a^2 \, , \label{eq:D_omega}
\end{align}
with $\tilde{\omega}_a = \sqrt{\omega_a^2 - 4 \eta_0 \omega_a}$ being the eigenfrequency of the closed system Hamiltonian in \cref{eq:H_sys_dispersive}.
The quantity $D(\omega)$ can be identified as the characteristic polynomial of the non-Hermitian matrix $\mathbf{A} - i \mathbf{\Gamma} / 2$.

We now linearize the coefficients $k_{\abs{n}\leq 1} (\omega)$ in $\epsilon$, which correspond taking the order zero in the expansion of $\tilde{k}(\omega)$. 
Specifically, for small $\epsilon$, we have that $\tilde{k}(\omega) \approx 2 i \gamma_a \omega / D(\omega) + O(\epsilon^2)$, which implies that the linear expansion of the output operators is
\begin{align}
    \tilde{c}_{\rm out}(\omega) =& \,\, \left( 1 - \frac{2 i \gamma_a \omega}{D(\omega)} \right) \tilde{c}_{\rm in}(\omega) \nonumber \\
    &+ \frac{4 i \epsilon \gamma_a \omega_a \sqrt{\omega (\omega + \omega_m)}}{D(\omega) D(\omega + \omega_m)} \tilde{c}_{\rm in}(\omega + \omega_m) \nonumber \\
    &+ \frac{4 i \epsilon \gamma_a \omega_a \sqrt{\omega (\omega_m - \omega)}}{D(\omega) D(\omega - \omega_m)} \tilde{c}_{\rm in}^\dagger(\omega_m - \omega)\,.
\end{align}
We notice that, as expected, the $k_{\abs{n} = 1} (\omega)$ coefficients are linear in $\epsilon$.

\section{Photon flux density}\label{sec:photon_flux_density}
\begin{figure}[t]
    \centering
    \includegraphics[width=\columnwidth]{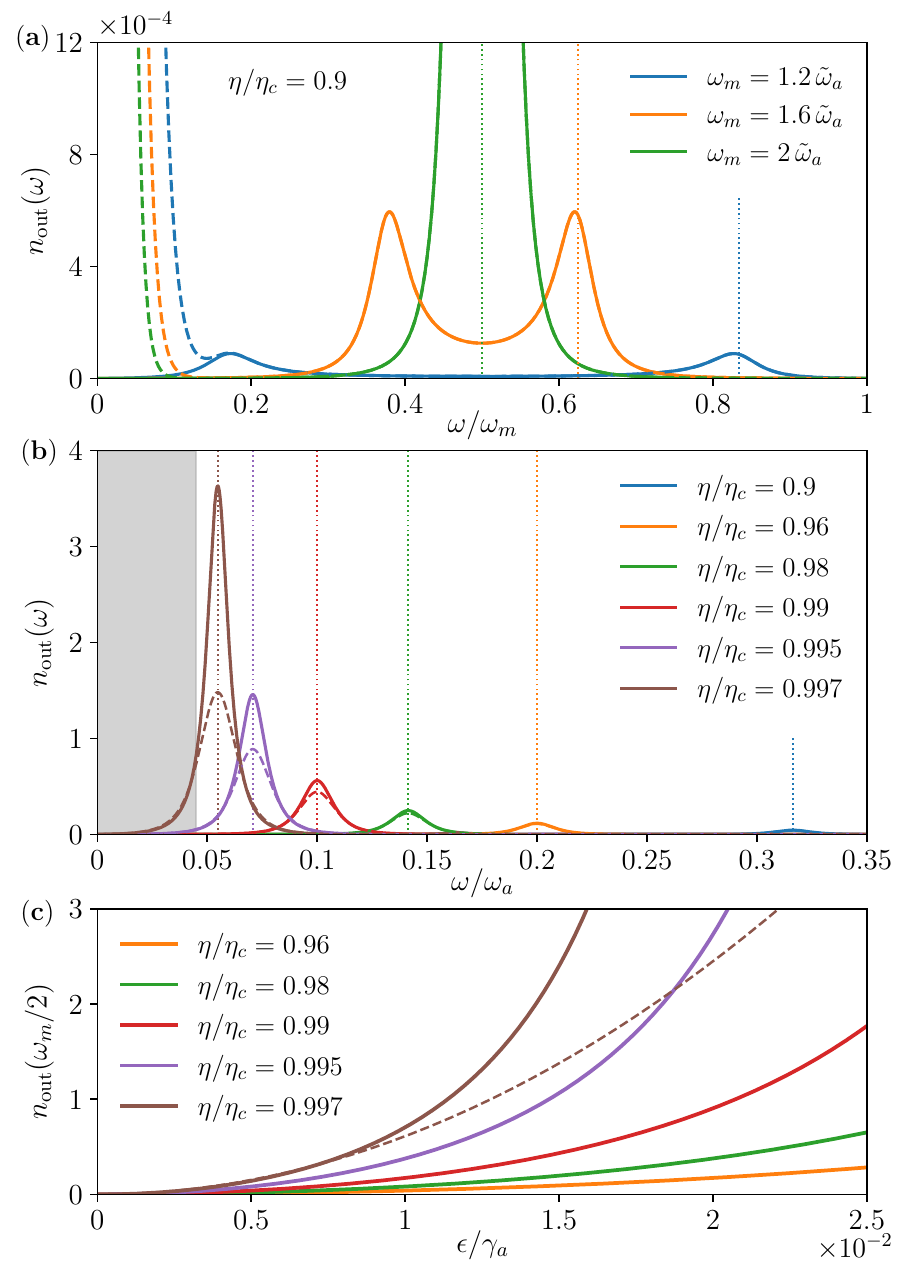}
    \caption{\textbf{Output photon flux density at non-resonant and resonant conditions.}
    \textbf{(a)} A double-peak structure emerges when the modulation frequency $\omega_m$ is detuned from twice the resonance frequency of the closed system $\tilde{\omega}_a$. Temperature effects (dashed curves) emerge as changes in the low-frequency region. \textbf{(b)} Comparison between the vacuum emission photon flux density calculated with \(N\) harmonics (solid lines) and the linear theory result (dashed curves) for different coupling strengths under resonant conditions. The linear approximation is accurate only for small modulation amplitudes and sufficiently far from criticality. The gray shaded region denotes parameter values that are inaccessible, since the chosen modulation amplitude would drive the system beyond the critical point.
    \textbf{(c)} Maximum output photon flux density, $n_{\mathrm{out}}(\omega)$, at $\omega_m/2$ as a function of the modulation amplitude $\epsilon$, normalized by the damping rate $\gamma_a$, for different coupling strengths $\eta/\eta_c$ (the dashed curve shows the linear theory result).
    The parameters used here are:  $\gamma_a/\omega_a =3 \times 10^{-2} $, (a) $\omega_\mathrm{th} / \omega_a = 0.005$, (a,b) $\epsilon/\gamma_a = (5/3)\times 10^{-2}$.}
    \label{fig:Vacuum_expectation_resonances}
\end{figure}

\begin{figure*}[t] 
    \centering
    \includegraphics[width=\textwidth]{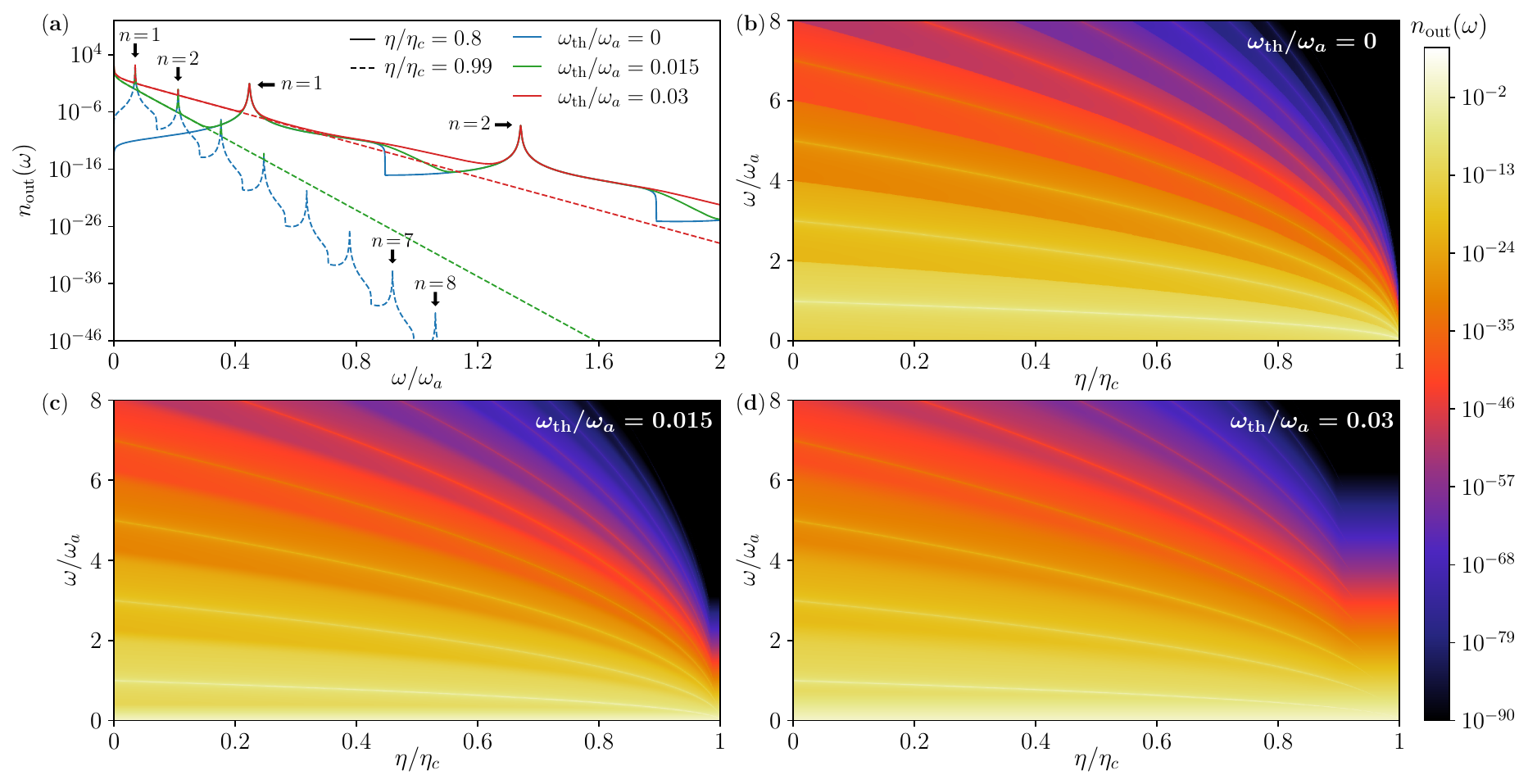}
    \caption{\textbf{Thermal vacuum emission.}
    \textbf{(a)} Output field photon flux density $n_{\mathrm{out}}(\omega)$ at resonance as a function of the mode frequency \(\omega\) for two relative coupling strengths, \(\eta/\eta_c = 0.8\) (solid) and \(\eta/\eta_c = 0.99\) (dotted), evaluated at different temperatures. An increasing number of harmonics contributes as the critical point is approached, as clearly visible at zero temperature ($\omega_\mathrm{th} = 0$, blue lines). \textbf{(b--d)} Colormaps at resonance of the vacuum emission photon flux density for increasing temperature.  The contributions of the various harmonics to the vacuum emission are distinctly separated; this separation fades as the temperature rises due to thermal broadening. The parameters used here are: $\gamma_a / \omega_a = 3 \times 10^{-3}$, $\epsilon / \gamma_a = 1/30$.}
    \label{fig:thermal_vacuum_emission}
\end{figure*}

To characterize the vacuum emission, we analyze the signal recorded by a photon intensity (power) detector, which defines the noise power spectrum. According to the Wiener-Khinchin theorem, for processes that are not strictly stationary, this quantity can be expressed in terms of the two-time correlation function of the output field as
\begin{equation}
\label{eq:photon_flux_density}
\begin{aligned}
n_{\rm out}(\omega) &= \!\frac{1}{2\pi} \lim_{T \to \infty} \frac{1}{T}\! \int_0^T \!\!\!\!\!\! dt \!\int_0^T \!\!\!\!\!\! dt' \, \langle c_{\mathrm{out}}^\dagger(t) \, c_{\mathrm{out}}(t') \rangle \, e^{-i\omega(t - t')} \\
&= \int_0^\infty \! d\omega^\prime \,
\big\langle c^\dagger_{\rm out}(\omega^\prime)\, c_{\rm out}(\omega) \big\rangle \,
\mathcal{F}_{\infty}(\omega^\prime-\omega)\, ,
\end{aligned}
\end{equation}
where \( n_{\rm out}(\omega) \) denotes the photon flux density, and \(\mathcal{F}_\infty(\alpha)\) acts as an effective \textit{frequency-selection filter} that suppresses all contributions with \(\alpha \neq 0\) and retains only the resonant terms with \(\alpha = 0\); its explicit definition, together with the derivation of this result, is provided in Appendix~\ref{app:derivation_emission_squeezing_spectra}.

Up to this point, all calculations have been performed at zero temperature. However the generalization to finite temperature is indeed possible by introducing $\bar{n}_{\mathrm{in}}(\omega) = \mathrm{Tr}\!\left[\rho\, c_{\mathrm{in}}^{\dagger}(\omega) c_{\mathrm{in}}(\omega)\right]$, which represents the thermal photon occupation of the input field mode with frequency $\omega$, given by $\bar{n}_{\mathrm{in}}(\omega) = \left[\exp\!\left( \omega/\omega_{th}\right) - 1\right]^{-1}$, where  $\omega_{th}=k_BT/\hbar$, in which $k_{\mathrm{B}}$ is the Boltzmann constant and $T$ is the temperature.

For illustrative purposes, we carry out the expansion of \eqref{eq:photon_flux_density} to first order ($N=1$) also incorporating thermal terms
\begin{align}\label{eq:thermal_1_photon_flux}
        n_{\rm out}(\omega) = &\,\abs{k_{0}(\omega)}^2 \bar{n}_{\rm in}(\omega) + \abs{k_{1}(\omega)}^2 \bar{n}_{\rm in}(\omega_m+\omega) \nonumber\\ &+\abs{k_{-1}(\omega)}^2 \bar{n}_{\rm in}(\omega_m-\omega)\theta(\omega_m-\omega) \nonumber \\
        &+\abs{k_{-1}(\omega)}^2 \theta(\omega_m-\omega) \, .
\end{align}
The first three terms in the foregoing expression are of thermal origin, whereas the fourth arises from the vacuum emission.
Starting from \eqref{eq:thermal_1_photon_flux} and specializing to zero temperature, we linearize, following the procedure outlined in \cref{Sec:model}, the only surviving term \(k_{-1}(\omega)\), obtaining
\begin{equation}
\label{eq:linearized_vac_emission}
n_{\rm out}(\omega) = \frac{16 \epsilon^2 \gamma_a^2 \omega_a^2 \omega (\omega_m - \omega)}{\abs{D(\omega)}^2 \abs{D(\omega-\omega_m)}^2}\, .
\end{equation}

In \figref{fig:Vacuum_expectation_resonances}(a) we present the output photon flux density as a function of the normalized frequency $\omega/\omega_m$ for different resonance frequencies. A double-peak structure appears when the modulation frequency is detuned from twice the resonance frequency of the closed system. Moreover, the peaks at $(\omega_m-\omega)$ and $\omega$ are displaced with respect to the resonances of the closed system (dotted vertical lines). By inspection of \eqref{eq:linearized_vac_emission}, this shift increases with the damping rate $\gamma_a$, which determines the imaginary part of the complex eigenvalues. Under resonant conditions (green line), the photon flux density is strongly enhanced. 
Temperature effects (dashed curves) emerge solely as modifications in the low-frequency region.

In \figref{fig:Vacuum_expectation_resonances}(b) we compare, under resonant conditions (i.e., $\omega_m = 2\,\tilde{\omega}_a$), and as the system approaches the critical point, the vacuum emission photon flux density computed by including the contributions of \(N\) higher harmonics (solid lines) with the prediction of the linearized theory (dashed curves). As expected, the linear description becomes inaccurate in the vicinity of criticality, confirming the necessity of a multi-harmonic higher-order description. The peaks of the photon flux density coincide with the resonance frequencies of the  closed system. For clarity, we include a gray shaded region and select values of \(\eta\) such that no peaks fall within this area, because the chosen modulation would otherwise drive the system beyond the critical point.

Figure~\ref{fig:Vacuum_expectation_resonances}(c) displays the maximum output photon flux density, $n_{\mathrm{out}}(\omega)$, evaluated at $\omega_m/2$ as a function of the normalized modulation amplitude $\epsilon/\gamma_a$, for different coupling strengths $\eta/\eta_c$. We observe that $n_{\mathrm{out}}(\omega)$ increases with the modulation amplitude: in the weak modulation region (small $\epsilon$) it exhibits the expected quadratic scaling, predicted by \eqref{eq:linearized_vac_emission} (see also the dashed curve), whereas for larger $\epsilon$ the growth becomes exponential. This crossover provides a signature of the non-perturbative character of the process under examination. 
It should be emphasized that we approach the quantum critical point from below the critical threshold, \(\eta \leq \eta_c = \omega_a/4\), and within the present framework we only consider modulations that do not take the system across the QPT. This restriction is purely theoretical rather than of the underlying physics, since crossing the critical point would require a different description \cite{Lamberto2026superradiantquantum}.

\begin{figure}
    \centering
    \includegraphics[width=\linewidth]{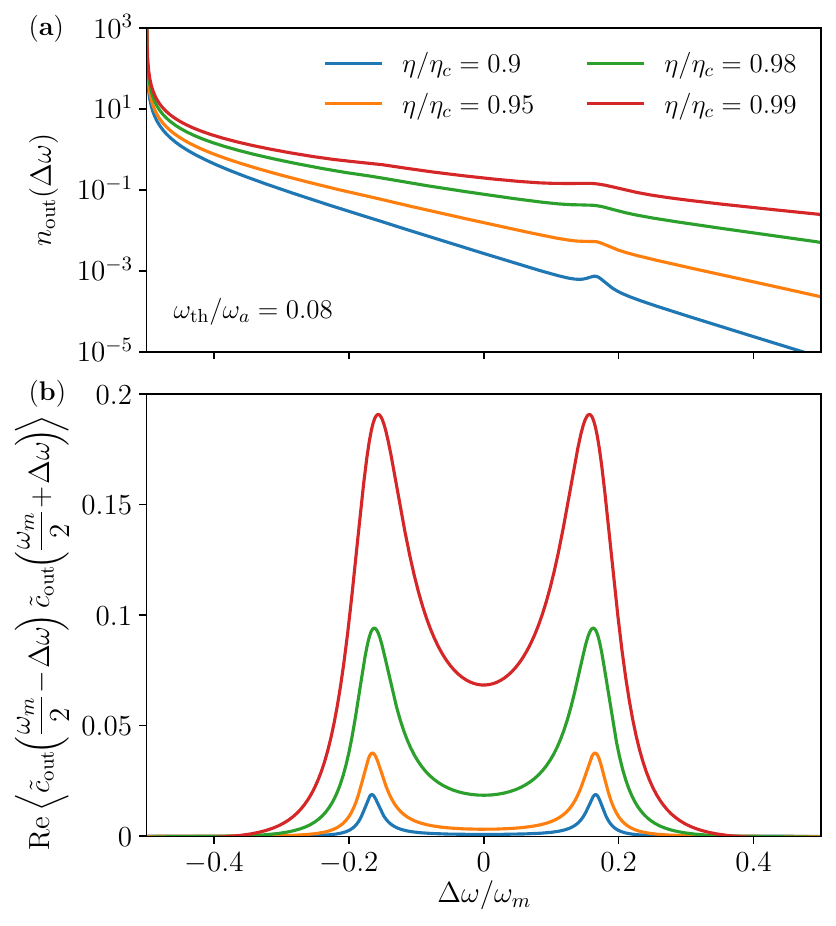}
    \caption{\textbf{Two-photon correlation.} \textbf{(a)} Output photon-flux density $n_{\mathrm{out}}(\omega)$ as a function of the normalized detuning $\Delta\omega/\omega_m$, showing that the harmonic response is dominated by thermal noise. \textbf{(b)} Two-photon correlation, which is strongly enhanced near the critical point and remains remarkably robust against thermal effects. $\omega_m = 3\,\tilde{\omega}_a$ is computed for each $\eta$. The parameters used here are: $\gamma_a / \omega_a = 0.03$, $\epsilon / \gamma_a = 1/30$, $\omega_{\rm th}/\omega_a=0.08$.}
    \label{fig:photons_corr}
\end{figure}

Figure~\ref{fig:thermal_vacuum_emission} reports the output photon flux density under resonant conditions, i.e., \(\omega_m = 2\,\tilde{\omega}_a\),  also showing the contribution of higher-order harmonics.
\figref{fig:thermal_vacuum_emission}(a) displays the resonant vacuum emission as a function of the normalized detection frequency \(\omega/\omega_a\) for different temperatures. 
The photon-flux density exhibits distinct harmonic peaks whose intensity decreases with harmonic order. Closer to the critical point, all peaks become significantly enhanced and these resonances cluster in the low-frequency regime, a direct manifestation of the mode softening of the lower polariton associated with the QPT (dashed lines in \figref{fig:thermal_vacuum_emission}(a), corresponding to $\eta / \eta_c = 0.99$).
Furthermore, the spectral lineshape is highly sensitive to thermal effects: at zero temperature, individual harmonics are well resolved, yielding characteristic step-like structures between successive peaks, whereas finite temperatures introduce significant resonance broadening.

The peaks structure  is determined by the energy conservation relation $\omega \,+\, \omega^\prime = n\,\omega_m$, where $n$ denotes the corresponding harmonic as shown in the \figref{fig:thermal_vacuum_emission}(a). An enhancement of the photon-flux density is observed whenever one of the two frequencies coincides with an integer multiple of the closed system resonance $\tilde{\omega}_a$. The enhancement is strongest, unsurprisingly, when both frequencies satisfy this condition simultaneously, which is precisely the case examined in Fig.~\ref{fig:thermal_vacuum_emission}. 
However, as the temperature rises, the thermal noise progressively increases and eventually masks the higher-order harmonics, making them no longer discernible in the spectrum.  We explicitly include the contributions of a large number of harmonics, reported on a logarithmic scale. Nevertheless, we recognize that such extremely small intensities are unlikely to be accessible in current experiments, at least for such low values of the modulation amplitude \(\left(\epsilon / \gamma_a = 1/30\right)\).
These trends are more evident in \figref{fig:thermal_vacuum_emission} (b--d): at zero temperature, the harmonic contributions are clearly resolved, and an increasing number of harmonics participates in the photon flux density approaching the critical point. By contrast, with rising temperature the distinction among harmonics progressively becomes less sharp, and near the QPT thermal noise fully masks  higher-order harmonics.

\begin{figure*}[t]
    \centering
    \includegraphics[width=\textwidth]{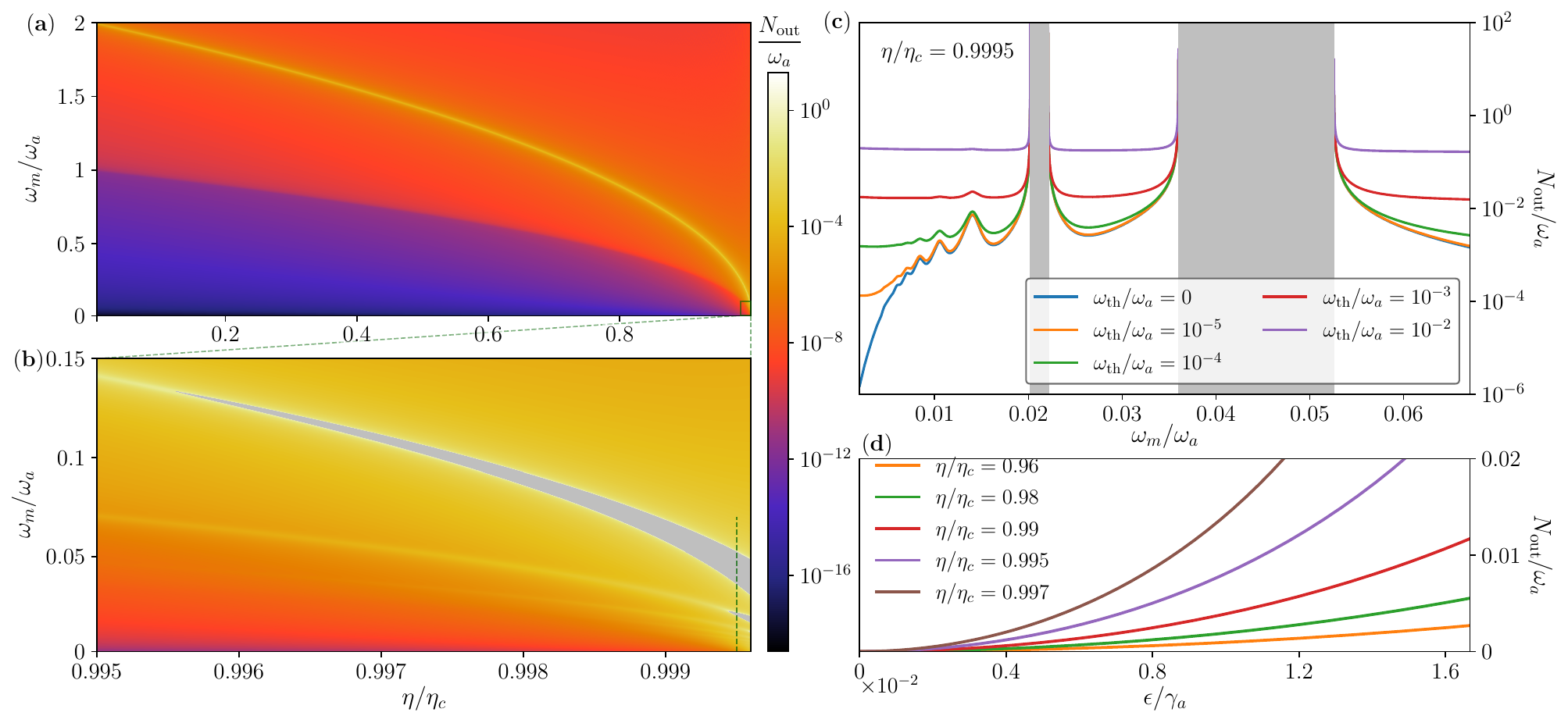}
    \caption{\textbf{Photon flux.}
    \textbf{(a)} Normalized output photon flux $N_\mathrm{out} / \omega_a$ as a function of the modulation frequency $\omega_m$ and the normalized coupling $\eta/\eta_c$. \textbf{(b)} Inset of panel (a) in the vicinity of the critical point. Upon approaching criticality, higher harmonics contribute appreciably to the emitted photon flux, and the onset of the instability region becomes visible (shaded grey area). \textbf{(c)} 
    Spectral cut of the photon flux in panel (b) (green dashed line) as a function of the modulation frequency $\omega_m$ for different temperatures. As $T$ increases, the thermal background progressively masks the contribution of higher harmonics, while the instability region (grey surface) remains unchanged.
    \textbf{(d)} Photon flux $N_{\rm out}/\omega_a$ at the first resonance $\omega_m = 2 \tilde{\omega}_a$ as a function of the modulation amplitude $\epsilon$, normalized by the damping rate $\gamma_a$, for different coupling strengths $\eta/\eta_c$. 
    The parameters used here are the same as in \cref{fig:thermal_vacuum_emission}.}
    \label{fig:photon_flux_spectrum}
\end{figure*}

The output field defined in \eqref{eq:out_in_definition} exhibits correlations between photons at distinct frequencies. A direct signature of these correlations is obtained by evaluating the expectation value \(\langle\tilde{c}_{\rm out}(\omega_m/2-\Delta\omega)\tilde{c}_{\rm out}(\omega_m/2+\Delta\omega)\rangle\) at two frequencies symmetrically distributed around $\omega_m/2$, where $\Delta\omega = \omega - \omega_m / 2$ is the detuning from the resonance condition. Physically, this quantity describes the simultaneous generation of photon pairs at $(\omega_m/2 - \Delta\omega)$ and $(\omega_m/2 + \Delta\omega)$, for $\Delta\omega < \omega_m/2$. Whereas the expectation value $\langle\tilde{c}_{\rm out}(\omega_m/2\pm\Delta\omega)\rangle$ vanishes, showing the absence of one-photon coherence, the two-photon correlation is different from zero and oscillates with the modulation frequency $\omega_m$.
The corresponding results are presented in \figref{fig:photons_corr}. While \figref{fig:photons_corr}(a) illustrates the output photon flux density as a function of the detuning $\Delta\omega$, revealing that the harmonic contribution is dominated by thermal noise, \figref{fig:photons_corr}(b) demonstrates a distinct behavior for the paired emission. Specifically, the correlation between the generated photon pairs not only increases as the system approaches the critical point, but also remains robust against thermal effects.

Experimentally, this correlation function can be probed through the Fourier transform of the voltage-noise signal, \(\big\langle V_{\rm out}^2(t)-\langle V_{\rm out}(t)\rangle^2\big\rangle\), with \(\langle V_{\rm out}(t)\rangle=0\) and $\tilde{V}_{\rm out}(\omega) \propto [\tilde{c}_{\rm out}(\omega)+\tilde{c}^\dagger_{\rm out}(\omega)]$. This procedure yields a contribution proportional to
\begin{equation} \label{eq:f_transf_V}\textit{F}\left[\big\langle V^2_{\rm out}(t)\big\rangle\right]\!(\omega_m)\!\propto  \!\!\int \!\!d\omega \; \mathrm{Re}\!\left\langle
\!\tilde{c}_{\rm out}\!\!\left(\frac{\omega_m}{2} \!-\!\omega\right)
\!\tilde{c}_{\rm out}\!\!\left(\frac{\omega_m}{2} \!+\!\omega\right)\!
\right\rangle ,
\end{equation}
the integrand of which is shown in \figref{fig:photons_corr}.
The measurement of this  quantity would make it possible to detect two-photon correlations even at finite temperature and arbitrarily close to the critical point.

\section{Photon Flux} \label{sec:photon_flux}

In \cref{sec:photon_flux_density} we have  provided a detailed characterization of vacuum emission in the vicinity of a QPT by examining the photon flux density $n_\mathrm{out} (\omega)$. In the following, we focus on the photon flux $N_\mathrm{out}$, which yields a quantitative estimate of the total number of photons emitted in response to a nonadiabatic temporal modulation applied near the critical point. Accordingly, we define
\begin{equation}
    N_{\rm out}=\int d\omega \, n_{\rm out}(\omega)\, ,
\end{equation}
where \(n_{\rm out}(\omega)\) is defined in \cref{eq:photon_flux_density}.

\figuref{fig:photon_flux_spectrum}(a) displays the emitted photon flux as a function of the modulation frequency $\omega_m$ and the normalized coupling $\eta/\eta_c$ as the system is tuned towards the critical point.
In \figref{fig:photon_flux_spectrum} (b), we analyze an inset of the preceding panel focusing on the immediate vicinity of the QPT, while keeping all parameters unchanged.
The conclusions inferred from the photon flux density in Fig.~\ref{fig:thermal_vacuum_emission} remain applicable: the harmonic contributions are sharply resolved, and an increasing number of harmonics (identified by the resonance condition $n\,\omega_m = 2\,\tilde{\omega}_a$) contributes to the photon flux as the critical point is approached. 

Notably, the  photon flux highlights an additional relevant feature, namely the onset of an instability region (shaded gray area) in close proximity to the QPT. In this region for sufficiently large modulation amplitudes the system develops an instability that gives rise to coherent parametric oscillations~\cite{Zerbe1995parametricquantumoscillator, WallsMilburn_book_input-output, DeLiberato2007vacuumradiation}. Once the instability threshold is exceeded, the Fourier-space solutions of Eq.~\ref{eq:QLE_input_extended} are not valid anymore, since the fields grow exponentially in time. The instability boundaries can be determined within a mean field treatment of the nonlinear differential equations for the field coordinates~\cite{Grimshaw2017nonlinear, Bender1999book}; a detailed derivation is deferred to the Appendix~\ref{sec:instability}. It should be noted that the values of $\eta/\eta_c$ used in \figref{fig:thermal_vacuum_emission} and \figref{fig:photon_flux_spectrum} do not extend into the instability region.



So far, we have discussed the photon flux at zero temperature as the system approaches the QPT. The extension to finite temperature is straightforward. To assess thermal effects, in \figref{fig:photon_flux_spectrum}(c) we consider a spectral cut at $\eta/\eta_c=0.9995$, i.e., strictly close to the critical point. We immediately observe that the instability region is insensitive to temperature, while the only effect of increasing $T$ is the enhancement of the thermal background, which masks the contribution of higher-order harmonics.

\figuref{fig:photon_flux_spectrum}(d) shows the normalized output photon flux, $N_{\rm out}/\omega_a$, versus the modulation amplitude, $\epsilon/\gamma_a$. As expected, the emitted photon flux grows with increasing modulation strength, and this growth becomes more pronounced when the system approaches criticality. Beyond illustrating this trend, \figref{fig:photon_flux_spectrum}(d) provides a quantitative estimate of experimentally relevant photon rates. Assuming, for example,  $\omega_a = 40~\mathrm{GHz}$, in line with the frequency scale near the critical point in the  experimental observation of the equilibrium SPT at low temperature~\cite{Kono2025observationspt}, one finds for  $\epsilon/\gamma_a = 0.01$ and $\eta/\eta_c=0.995$ (red line), an output photon flux of
\(
N_{\rm out} \simeq 4\times10^8 \, \mathrm{s^{-1}}.
\)
This demonstrates not only that the proximity to the critical point can significantly enhance the flux of detectable emitted photons, but also underscores the realistic experimental feasibility of observing these emitted photons.

\section{Squeezing} \label{sec:squeezing}

In contrast to non-equilibrium QPTs, quantum phenomena (such as long-range correlations and  ground state squeezing) remain inaccessible to direct experimental observation in equilibrium QPTs, as they involve virtual rather than real excitations \cite{Lamberto2026superradiantquantum}. 

The generation of real particles from the quantum vacuum arises from Heisenberg’s uncertainty principle, whereby vacuum fluctuations can be promoted to observable excitations under appropriate conditions. In the ultrastrong coupling regime, the ground state of light-matter systems hosts virtual correlated excitations, such as photon-matter pairs in a squeezed vacuum. However, these excitations are bound within the system and cannot radiate \cite{FriskKockum2019ultrastrongNRP,FornDiaz2019ultrastrongRMP}.

Nonadiabatic modulation of the quantum vacuum, for instance through rapid time variation of the light-matter coupling, can release such excitations as detectable quantum vacuum radiation, analogous to the Moore effect, which is also known as the dynamical Casimir effect \cite{Moore1970dce,dodonov2020fifty}. Furthermore, it has been shown that, in the vicinity of the critical point, ground state quantum squeezing (a redistribution of quantum uncertainty between complementary observables) increases, reaching perfect squeezing precisely at the critical point.

However, we demonstrated in our previous work \cite{Lamberto2026superradiantquantum} that ground state squeezing cannot be directly measured with current techniques, including electro-optic sampling of vacuum radiation \cite{Benea2019electric, Riek2015direct, Lindel2020quantumvacuumdetection}. Here we examine whether the nonadiabatic modulation of a Hamiltonian parameter driving the QPT can convert virtual ground state correlations into real and detectable properties of the output field, thus making quantum vacuum squeezing observable.

\begin{figure*}[t]
    \centering
    \includegraphics[width=\textwidth]{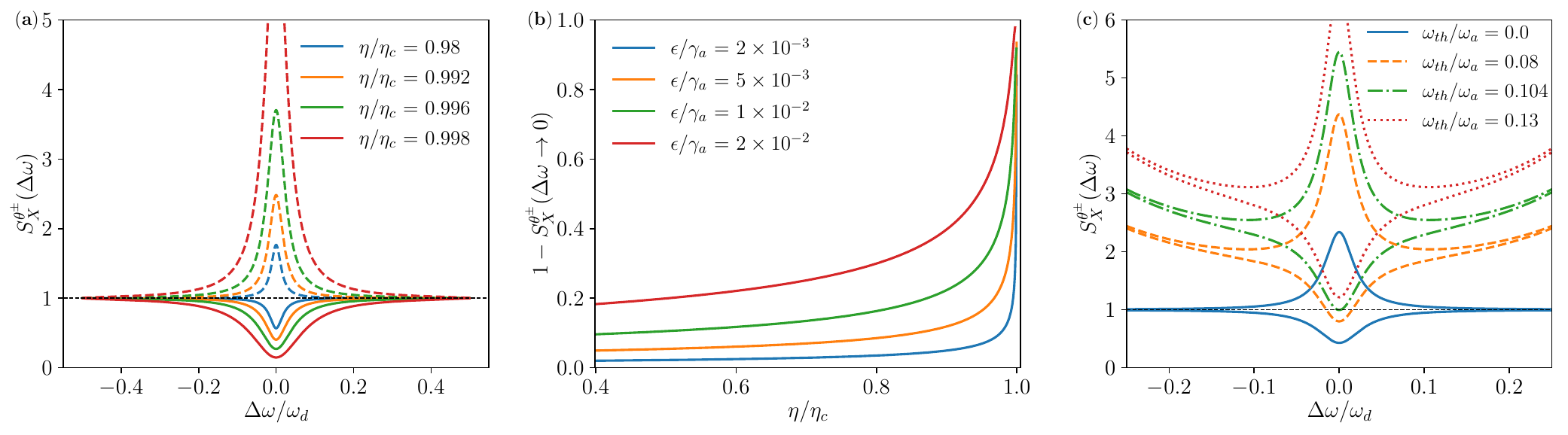}
    \caption{\textbf{Squeezing Spectrum.} 
\textbf{(a)} Squeezing spectrum $S^{\theta_\mathrm{opt}}_X(\Delta\omega)$ at zero temperature as a function of detuning from the resonant condition $\Delta\omega$, evaluated at the optimal angle $\theta_\mathrm{opt}$. Approaching the critical point, the squeezing approaches its ideal limit. 
\textbf{(b)} Fraction of squeezing versus the normalized coupling strength $\eta / \eta_c$ for four modulation amplitudes, showing that near the QPT almost $100\%$ squeezing is achieved. 
\textbf{(c)} Squeezing spectrum at finite temperature as a function of detuning, illustrating the degradation of quantum squeezing with increasing temperature (green line). 
The parameters used here are: $\gamma_a/\omega_a = 10^{-2}$; (a) $\epsilon/\gamma_a = 10^{-2}$; (c) $\epsilon/\gamma_a = 10^{-2}$, $\eta / \eta_c = 0.99$.}
    \label{fig:squeezing_spectrum}
\end{figure*}

\subsection{Squeezing Spectrum} \label{subsec:squeezing_spectrum}

A nonclassical manifestation of the pairwise photon correlations arising in the fields described by \eqref{eq:out_in_definition} is quadrature squeezing \cite{Caves1985quadrature}. This effect can be quantitatively characterized through the squeezing spectrum \cite{CollettWalls1985squeezingspectra, WallsMilburn_book_input-output,Scully_book,carmichael2007statistical,agarwal2012quantum}, which specifies quadrature squeezing as a function of frequency. 
The field quadratures are formally defined by the relation
\begin{equation}
    X^{\theta}_{\rm out}(t)=c_{\rm out}(t)\, e^{i\theta} + c^\dagger_{\rm out}(t)\, e^{-i\theta} \, .
\end{equation}
Experimentally, the quadratures of a continuous multimode field are accessed via homodyne detection, wherein the signal field is combined with a local oscillator (LO) on a balanced beam splitter. Assuming the LO to be in a large-amplitude coherent state with well-defined frequency \(\Omega\) and phase \(\theta\), the subsequent intensity measurement of the output field yields direct information about the quadratures of the signal field 

\begin{equation}
    X^{\theta}_{\rm out}(t)=c_{\rm out}(t)\, e^{i(\theta + \Omega  t)} + c^\dagger_{\rm out}(t)\, e^{-i(\theta + \Omega t)} \, .
\end{equation}

The noise-power spectrum of the output field directly yields the squeezing spectrum of the signal field in the rotating frame at frequency \(\Omega\), namely \cite{Scully_book, Johansson2010dcecircuitPRA}
\begin{align} \label{eq:squeezing_spectrum}
    S^\theta_X(\Delta\omega)&= \, 1+\frac{1}{2\pi}\,  \lim_{T\rightarrow\infty}\frac{1}{T}\int_0^T \!\! dt \nonumber \\ &\quad \qquad \int_0^T \!\! dt^\prime \langle : \! X^{\theta}_{\rm out}(t) X^{\theta}_{\rm out}(t^\prime) \! : \rangle \, e^{-i \Delta\omega(t-t^\prime)}  \, ,
\end{align}
where $\langle : \, : \rangle$ denotes the normally ordered expectation value, and the squeezing spectrum has been normalized such that $S_{X}^{\theta}=1$ for the unsqueezed vacuum and $S_{X}^{\theta}=0$ for maximal squeezing. The frequency $\Delta\omega$ measured after mixing with the LO is related to the signal field frequency by $\omega = \Omega + \Delta \omega$. Note that in \eqref{eq:squeezing_spectrum}, whose derivation is presented in the Appendix~\ref{app:derivation_emission_squeezing_spectra}, we employ the general formula for power spectra, since the process under consideration is not strictly stationary.

The squeezing spectrum, both at zero and finite temperature, is reported in \figref{fig:squeezing_spectrum}. Panel \ref{fig:squeezing_spectrum}(a) illustrates the squeezing spectrum at zero temperature as a function of the detuning from the resonant condition, $\Delta\omega = \omega - \omega_m / 2$, for different values of $\eta$ near the critical point. As the system approaches the critical point, i.e., $\eta / \eta_c \to 1$, the squeezing tends to become perfect. 

\figuref{fig:squeezing_spectrum}(b) presents the same analysis from a different perspective, showing the percentage of squeezing, obtained from the maximum of the squeezing spectrum calculated in the limit  $\Delta\omega \rightarrow 0$, as a function of the coupling strength for different modulation amplitudes. In this case, it is evident that, near the QPT, almost $100\%$ squeezing is achieved. This is a key result, as it demonstrates that perfect ground-state squeezing predicted theoretically \cite{Brandes2003DickeqptPRE, Hayashida2023squeezingDicke} can be converted, via nonadiabatic modulation, into an experimentally observable phenomenon. Just as vacuum emission in USC systems can be interpreted as the ability of a modulation to convert virtual photons into real ones, we interpret the observation of this squeezing, which tends to become perfect near criticality, as evidence that the modulation can convert virtual ground state squeezing into a real, observable one.

Finally, \figref{fig:squeezing_spectrum}(c) displays the squeezing spectrum at finite temperature. As expected, thermal effects degrade quantum squeezing, since increasing the temperature leads to a minimum squeezing value exceeding unity. However, we observe that the imbalance between the quadrature noises is maintained. This effect will be further analyzed in \cref{sec:entanglement}. All these panels have been calculated for an optimal angle $\theta_\mathrm{opt} \approx \pi / 4$ which maximizes the squeezing, whose precise definition will be formulated in the next paragraph.

\begin{figure*}
    \centering
    \includegraphics[width=\linewidth]{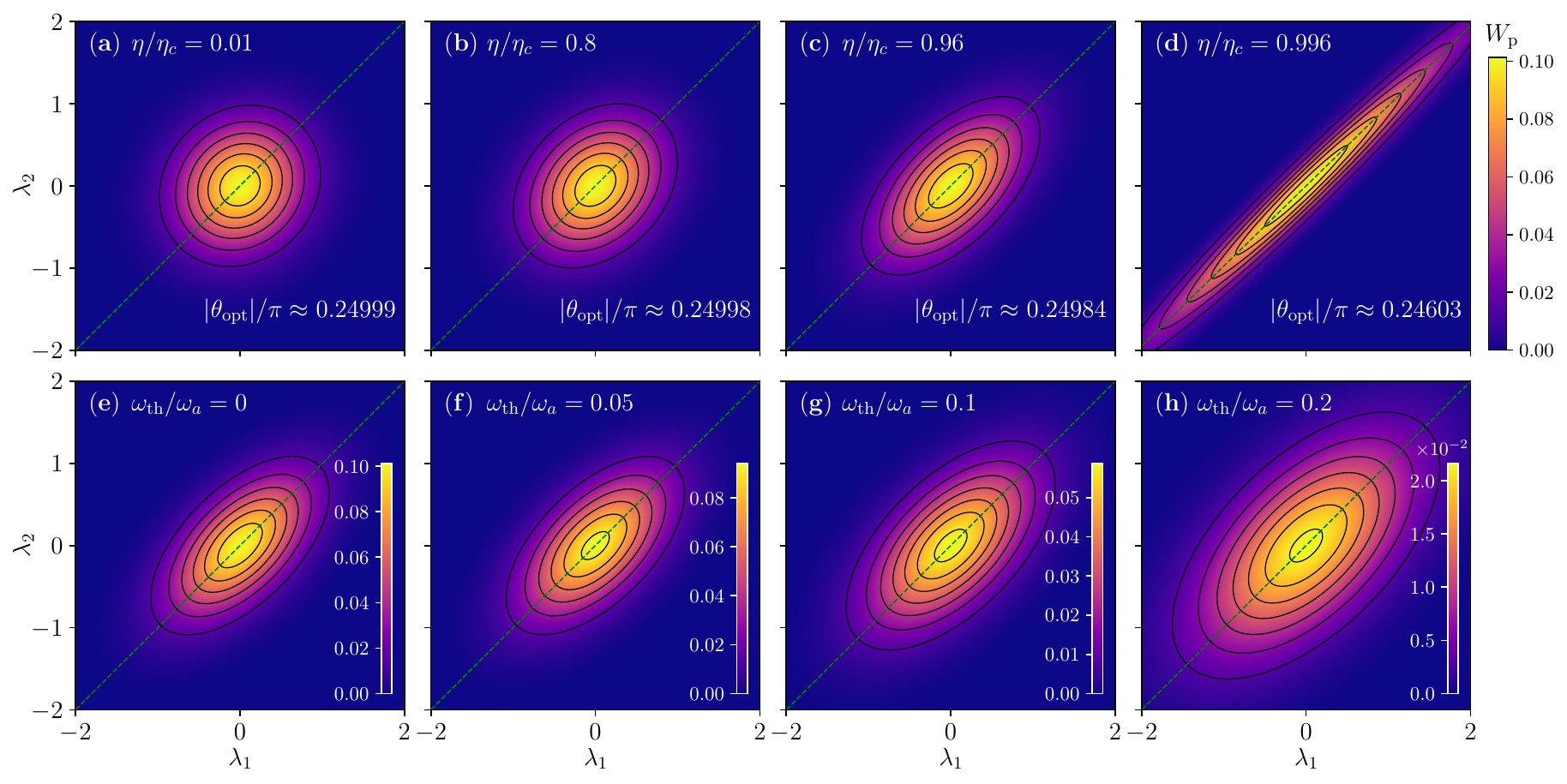}
    \caption{\textbf{Wigner functions of the output field.} \textbf{(a--d)} Wigner functions at zero temperature for the ground state of the output field, for increasing coupling strengths $\eta/\eta_c$. We observe the progressive squeezing of the output state, which tends to be perfect at the critical point. \textbf{(e--h)} Wigner functions for several thermal states for $\eta / \eta_c = 0.96$. As the temperature increases, the axes of the Wigner functions expand, while their ratio remains constant. Parameters: $\gamma_a/\omega_a =3 \times 10^{-2} $, $\epsilon/\gamma_a = 1.67\times 10^{-2}$.}
    \label{fig:Wigner_function}
\end{figure*}

\subsection{Wigner Representation} \label{subsec:Wigner}

It is also illustrative to compute the Wigner function of the output fields, both for the ground (for $T=0$) and for a thermal state ($T \neq 0$). 
As shown below, its behavior further confirms the presence of two-mode squeezing. In particular, since both the ground state of the Hamiltonian and thermal states are Gaussian, the Wigner function $W(\mathbf{r})$ is fully characterized by the first and second moments of the quadrature operators, namely the vector of the expectation values and the covariance matrix $V$, respectively, given by \cite{Johansson2013nonclassical}
\begin{align} 
    \langle \mathbf{r} \rangle &= \mathbf{0} \\
    \!\!\! V_{\alpha\beta} &=\frac{1}{2}\langle r_\alpha r_\beta \!+\! r_\beta r_\alpha \rangle \!- \!\langle r_\alpha \rangle \langle r_\beta \rangle = \frac{1}{2}\langle r_\alpha r_\beta \!+\! r_\beta r_\alpha \rangle \,, \label{eq:covariance mat}
\end{align}
where \(\mathbf{r}=(q_-,p_-,q_+,p_+)^T\) denotes the vector of quadratures, defined as
\begin{equation}\label{eq:quadratures}
    q_\pm = \frac{1}{\sqrt{2}}\left(b_\pm + b_\pm^\dagger\right) \, , \quad
    p_\pm = -\frac{i}{\sqrt{2}}\left(b_\pm - b_\pm^\dagger\right) \, ,
\end{equation}
in which \(b_+=c_{\rm out}(\omega)\) and \(b_-=c_{\rm out}(2\Omega-\omega)\). 

Although the quadratures $q_\pm$ and $p_\pm$ could in principle be chosen more generally, we adopt the canonical definitions in \eqref{eq:quadratures}, which mimic the position- and momentum-like operators of bosonic modes and corresponds to the natural zero-phase reference for the Wigner function. 
For such Gaussian states, the Wigner function can be written as \cite{Weedbrook2012gaussianRMP}
\begin{equation}
    W (\mathbf{r}) = \frac{1}{(2 \pi)^2 \sqrt{\det V}} \, \exp\left( -\frac{1}{2} \mathbf{r}^T \, V^{-1} \, \mathbf{r} \right) \, .
\end{equation}

In \cref{fig:Wigner_function}(a--d), we plot the Wigner function of the ground state of the output field at $T=0$ for increasing coupling strengths, approaching the critical point. Specifically, we show the projection of the Wigner function onto the two reference axes $\lambda_1$ and $\lambda_2$, $W_\mathrm{p}(\lambda_1, \lambda_2)$, which correspond to the zero-phase squeezing coordinates $(q_- + q_+)$ and $(p_- + p_+)$, respectively. As the critical point is approached, we observe a progressive squeezing of the ground state of the output field. 
Direct inspection of the covariance matrix (see Appendix~\ref{app:covariance_matrix}) reveals that the state corresponds to a two-mode squeezed vacuum with squeezing angle $2\theta_\mathrm{opt} \approx \pi / 2$. The deviation from the value $\theta_\mathrm{opt} = \pi / 4$ arises from the coupling between the system and the external environment, and this shift progressively increases as $\eta \to \eta_c$, as reported in the corresponding panels. In particular, $\theta_\mathrm{opt}$ can be calculated analytically and it is given by the phase of the expectation value $\langle b_- b_+ + b_+ b_- \rangle$, namely $\arctan(-\operatorname{Im}\langle b_- b_+ + b_+ b_- \rangle / \operatorname{Re}\langle b_- b_+ + b_+ b_- \rangle)$.

\figuref{fig:Wigner_function}(e--h) shows the behavior of the Wigner function for thermal states at different temperatures and fixed $\eta / \eta_c = 0.96$. As the temperature increases, the two axes of the Wigner function progressively expand (further confirming the trend observed in \figref{fig:squeezing_spectrum}~(c)), while their ratio remains constant.

\section{Quantum properties: Nonclassicality and entanglement} \label{sec:entanglement}

The ground state of light-matter systems in the USC inherently hosts virtual quantum-correlated excitations. It has been shown \cite{FriskKockum2019ultrastrongNRP,FornDiaz2019ultrastrongRMP} that these bound photons can be released when the quantum vacuum is nonadiabatically modulated in time and we demonstrated in \cref{sec:squeezing} that a fundamental quantum property, i.e., squeezing, is  enhanced in the vicinity of a quantum critical point. Building upon these findings, we now aim to apply established nonclassicality criteria and evaluate the entanglement of the resulting field states, thereby providing further evidence of their quantum nature.
We consider an operator \( f \), defined as a function of the creation and annihilation operators. For the Hermitian operator \( f^{\dagger} f \), it can be shown~\cite{Miranowicz2010testing}, within the Glauber--Sudarshan \( P \)-function formalism, that any classical field state satisfies \( \langle :\!f^{\dagger} f\!: \rangle \ge 0 \), where classicality requires a non-negative \( P \)-function. For the two-mode quadrature-squeezed states generated by the system under study, a natural choice of \( f \) is

\begin{align}
    f_\theta = \,&e^{i(\theta+\pi/4)}c_{\rm {out}}(2\Omega-\omega) + e^{-i(\theta+\pi/4)}c^\dagger_{\rm {out}}(2\Omega-\omega)\nonumber\\ 
    &  + i\left(e^{i(\theta+\pi/4)}c_{\rm {out}}(\omega) - e^{-i(\theta+\pi/4)}c^\dagger_{\rm {out}}(\omega)\right) \, ,
\end{align}
where \( \theta \) denotes the angle defining the principal squeezing axis. This choice of \( f_{\theta} \) is also experimentally convenient, as \( \langle :\!f_{\theta}^{\dagger} f^{\vphantom{\dagger}}_{\theta}\!: \rangle \) can be directly determined from measurable quadrature correlations. Within this definition we obtain
\begin{align} \label{eq:nonclassicality}
    \langle :\!f_{\theta}^{\dagger} f^{\vphantom{\dagger}}_{\theta}\!: \rangle \!=& 2 \left(\!\langle c_{\rm out}^\dagger(2\Omega\!-\!\omega)c_{\rm out}(2\Omega\!-\!\omega)\rangle \!+\! \langle c_{\rm out}^\dagger(\omega)c_{\rm out}(\omega)\rangle \! \right) \nonumber \\
    &- e^{2i\theta}\langle c_{\rm out}(2\Omega-\omega)c_{\rm out}(\omega)\rangle \nonumber\\
    &- e^{-2i\theta}\langle c_{\rm out}^\dagger(\omega)c_{\rm out}^\dagger(2\Omega-\omega)\rangle \nonumber \\
    &- e^{2i\theta}\langle c_{\rm out}(\omega) c_{\rm out}(2\Omega-\omega)\rangle \nonumber\\
    &- e^{-2i\theta}\langle c_{\rm out}^\dagger(2\Omega-\omega)c_{\rm out}^\dagger(\omega)\rangle \, .
\end{align}

Equation~ (\ref{eq:nonclassicality}) can be used to provide clear evidence that the emitted radiation exhibits nonclassical behavior. The quantum nature of this radiation originates from the entanglement between individual photon pairs, so to quantify the degree of entanglement between two collective modes whose frequencies sum to the modulation frequency, we evaluate the logarithmic negativity \( \mathcal{N} \), a widely employed entanglement measure in quantum optics \cite{Plenio2005logarithmicnegativity}. It takes positive values for entangled states and can be computed directly from the covariance matrix of the two selected modes (see \eqref{eq:covariance mat}) as
\begin{equation}
    \mathcal{N}=\rm{max}[0,-\rm{log(2\nu_-)}],
\end{equation}
where we defined 
\begin{align}
    \nu_- = & \left[\sigma/2-\left(\sigma^2 - 4\, \rm{det}V \right)^{1/2}/2 \right]^{1/2} \, , \\
 \sigma = & \, \langle b_- b_-^\dagger + b_-^\dagger b_- \rangle^2 + \langle b_+ b_+^\dagger + b_+^\dagger b_+ \rangle^2 \nonumber \\
 & -2  \abs{\langle\left(b_- b_+ + b_+ b_-\right)\rangle}^2 \, .
\end{align}

\begin{figure}[t]
    \centering
    \includegraphics[width=\columnwidth]{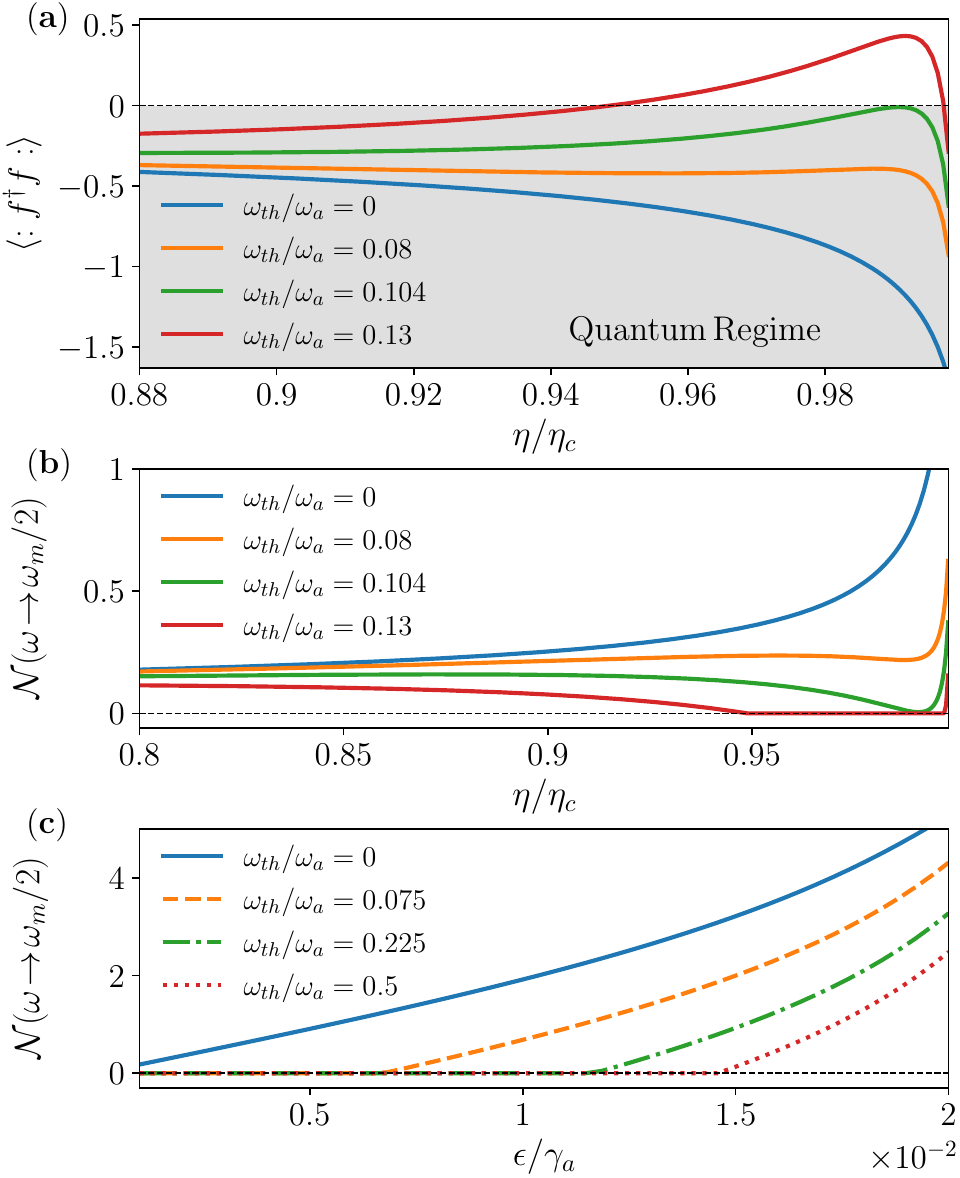}
    \caption{\textbf{Nonclassicality and Entanglement.}
    \textbf{(a-b)} Plots of (a) the quantum-classical indicator  $\langle :\! f^{\dagger} f \!:\rangle$ and (b) the logarithmic negativity $\mathcal{N}$ as functions of the relative coupling strength $\eta / \eta_c$, evaluated at the optimal angle $\theta$ and under resonant conditions.  For the chosen set of parameters, classical behaviour emerges away from the QPT when the temperature exceeds a critical threshold (green curve). However, independently of the temperature, both nonclassicality (a) and entanglement (b) are restored in the immediate vicinity of the critical point. \textbf{(c)} Maximum logarithmic negativity \(\mathcal{N}\) at \(\omega_m/2\) as a function of the modulation amplitude \(\epsilon\). The onset of a nonzero \(\mathcal{N}\) shifts toward larger modulation amplitudes with increasing temperature. Parameters:  \(\gamma_a/\omega_a = 10^{-2}\); (a,b) \(\epsilon/\gamma_a = 10^{-2}\); (c) \(\eta/\eta_c = 0.998\).}
    \label{fig:entanglement}
\end{figure}

In \figref{fig:entanglement} we show the nonclassical properties of the vacuum emission and the entanglement between pairs of modes whose frequencies sum to the modulation frequency, as the system approaches the critical point, both at zero and at finite temperatures. In \figref{fig:entanglement}(a) we plot the quantum-classical indicator \(\langle :\!f^{\dagger} f\!:\rangle\) as a function of the relative coupling strength \(\eta/\eta_c\), evaluated at the optimal angle \(\theta\) and under resonant conditions. 

For the set of parameters considered here, we find that at sufficiently low temperatures the observable is negative, \(\langle :\! f^{\dagger} f \!:\rangle < 0\), which signals a nonclassical state of the output field. As the temperature is increased and the system moves towards the quantum critical point, the influence of the thermal input field becomes more pronounced and eventually drives the indicator to positive values, \(\langle :\! f^{\dagger} f \!:\rangle > 0\), indicating that thermal noise dominates over quantum  fluctuations. Remarkably, in the immediate vicinity of the QPT the system re-enters the quantum regime. This  behavior is consistent with the parametric nature of the underlying process: in a parametric mechanism there is an intrinsic competition between photon pair production, which enhances quantum  correlations, and decoherence induced by temperature \cite{Galve2010bringing}. 

\figuref{fig:entanglement}(b) illustrates the corresponding behavior of the logarithmic negativity \(\mathcal{N}\) in resonant conditions as a function of \(\eta/\eta_c\). At zero temperature, the entanglement between the two matching output modes increases when the system approaches the QPT. As the temperature rises, thermal noise progressively degrades quantum correlations and, for the parameters adopted here,  when the temperature exceeds a critical threshold (green curve) the two modes are no longer entangled. 

Note that this critical threshold coincides with the value at which in \figref{fig:squeezing_spectrum}(c) quantum squeezing, i.e., squeezing below unity, is no longer observed.
Entanglement is restored only in the immediate vicinity of the QPT, which identifies the region in which the quantum correlations generated by the vacuum  emission dominate over thermal noise. This restoration should not be regarded as surprising: it is another manifestation of the same competition between pair creation and thermal decoherence mentioned above. The fact that both the nonclassicality indicator in \figref{fig:squeezing_spectrum}(a) and the entanglement measure in \figref{fig:squeezing_spectrum}(b) recover their quantum nature near the critical point shows that, in this regime, photon pair production overwhelms the effect of thermal decoherence, despite the softening of the mode.

\figuref{fig:entanglement}(c) displays the maximum value of the logarithmic negativity  \(\mathcal{N}\), evaluated in the \(\omega \rightarrow \omega_m/2\) limit, as a function of the modulation amplitude \(\epsilon\). 
We observe that the onset of a nonzero \(\mathcal{N}\) shifts towards larger modulation amplitudes as the temperature is increased. Beyond this onset, the logarithmic negativity increases with the modulation amplitude, showing that sufficiently strong modulation is able to generate entanglement even in the presence  of finite temperature.

\section{Conclusions}
\label{sec:conclusions}

Virtual ground-state excitations and quantum correlations can increase significantly near a quantum critical point \cite{Lambert2004entanglementDicke, Hayashida2023squeezingDicke}. We have demonstrated that a nonadiabatic modulation of a Hamiltonian parameter can efficiently convert these virtual excitations into real, detectable photons.
Proximity to criticality strongly amplifies the emitted photon flux and enhances the nonclassical character of the radiation, leading to measurable squeezing and entanglement even in regimes where thermal noise is significant. 
In particular, as shown in \cref{fig:photons_corr}, the generated photon pairs remain strongly correlated in the proximity of the critical point even at nonzero temperature. 
Experimentally, this correlation can be probed through the Fourier transform of the voltage-noise signal, \(\big\langle V_{\rm out}^2(t)-\langle V_{\rm out}(t)\rangle^2\big\rangle\), as described by \eqref{eq:f_transf_V}, thereby offering a potential route toward the development of critically enhanced sensors.

We further demonstrate that higher-order resonance processes become increasingly relevant close to the critical point and must be included to obtain quantitatively accurate predictions. The exact quantum Langevin framework employed here provides a systematic, analytical, and experimentally relevant description of these effects at zero and finite temperature, including the region in the proximity of the critical point where the Born-Markov approximation (underlying the standard master-equation approach) breaks down.
In addition to yielding concrete predictions for spectral observables, quadrature statistics, and entanglement measures that can directly guide experiments, this formalism provides a versatile tool for exploring vacuum emission in general material systems collectively coupled to electromagnetic resonators \cite{Roche2025}.

The modulation protocol described here can be easily implemented in the recently observed magnonic Dicke SPT \cite{Kono2025observationspt} by superimposing a controlled time-dependent component onto the static magnetic field. Specifically, a periodic Zeeman modulation of the form $B(t)=B_0+\delta B\cos (\omega _{\mathrm{d}}t)$ can convert the system’s virtual ground-state excitations into real photons, with a particularly intense emission near the critical point. Demonstrating vacuum radiation in such systems would provide a concrete experimental test of the criticality-enhanced conversion of virtual correlations into observable quantum resources.

Looking ahead, we plan to extend the present analysis into the broken-symmetry phase beyond the critical point in order to investigate how the modulation interacts with the system condensates. Mapping the dependence of emission spectra, squeezing, and entanglement on the order parameter could clarify whether the critical amplification persists, saturates, or gives rise to qualitatively different dynamical phenomena in the superradiant phase.

It would also be interesting to extend these results to finite spin ensembles. Studying systems of varying size wouls make it possible to quantify how the critical enhancement of vacuum radiation evolves from few-body to many-body, and hence to identify the minimal system size required to observe the predicted amplification and nonclassical signatures.

\appendix

\section{Derivation of the QLEs and the Input Fields}
\label{app:derivation_langevin_input}

The purpose of this Appendix is to introduce the formalism of QLEs that will be employed throughout the paper, following the procedure outlined in Ref.~\cite{GardinerZoller_book}. We assume that the baths can be modeled as an infinite set of independent harmonic oscillators, a description that becomes exact, for instance, for electromagnetic environments.

The QLEs are obtained by solving the equations of motion for the bath degrees of freedom in terms of the system variables and substituting the resulting expressions into the system Heisenberg equations, namely
\begin{align} \label{app:QLE_generic}
    \dot{O} &= \frac{i}{\hbar}\left[H_\mathrm{sys},O\right] \nonumber \\
    &-\! \frac{i}{2\hbar} \! \left[\left[X_a,O\right] \, , \; \xi(t) - \! \int_{t_0}^t \! \dot{X}_a(t') f(t-t') \, dt' \right. \nonumber \\
    & \,\qquad\qquad\qquad\qquad\left. - f(t-t_0)X_a(t_0) \vphantom{\int} \right]_+ \, ,
\end{align}
where $X_a$ is the system coordinate, responsible of the coupling between the system and the bath, and \(O(t)\) is an arbitrary system operator. Moreover, \([\dots,\dots]_+\) denotes the anticommutator, while \(f(t)=\sum_n k_{n}\cos(\omega_{n}t)\) plays the role of a memory function, and
\begin{equation} \label{eq:xi_ab}
    \xi(t)= \sum_n \sqrt{\frac{\hbar k_{n} \omega_{n}}{2}}
    \left( c_{n}^\dagger(t_0)\, e^{i\omega_{n}(t-t_0)} + \mathrm{h.c.} \right)
\end{equation}
is the bath noise operator.
By specializing \cref{app:QLE_generic} to our system's Hamiltonian and taking $O(t) = a(t)$, we obtain
\begin{align} \label{app:QLE_a}
    \dot{a}(t) = & -i\omega_a a(t) + 2i \eta(t) \left[ a(t) + a^\dagger (t) \right] +\frac{i}{\sqrt{2\hbar\omega_a}} \xi(t) \nonumber \\
    & -\frac{i}{2\omega_a}\int_{t_0}^t \! f(t-t')[\dot{a}^\dagger(t')+\dot{a}(t')] dt' \, . 
\end{align} 

We now analyze \cref{app:QLE_a} and its Hermitian conjugate in the frequency domain, assuming that the initial time lies in the remote past, i.e., \(t_0\to -\infty\). Under this assumption, one obtains the QLEs for the input fields. Conversely, by considering the limit $t_0 \to + \infty$, the QLEs for the output fields can be derived. Either of these limits, together with the assumption that \(f(t)\) is strongly localized near the origin of its argument, allows the last term in \cref{app:QLE_generic} to be neglected. 

We define the decay-rate function \(\gamma(\omega)\) as the Fourier transform of \(\theta(t) f(t)\), where \(\theta(t)\) is the Heaviside step function introduced to enforce causality. In general \(\gamma(\omega)\) is complex-valued, however, since \(f(t)\) is real-valued, it must satisfy \(\gamma^*(\omega)=\gamma(-\omega)\). Without loss of generality, in the present work we assume \(\gamma(\omega)\) to be real.
Under these assumptions we specialize the system Hamiltonian in \eqref{eq:H_sys_dispersive} to the case of a periodic modulation of the coupling strength at frequency $\omega_m$, namely \(\eta(t)=\eta_0+\epsilon \cos(\omega_m t)\,\). The QLEs can then be written in the compact form given in \cref{eq:QLE_input_extended} of the main text,
\begin{align}
    - i \omega\, \tilde{\mathbf{v}}(\omega)
    =& - i \left(\mathbf{A} - \frac{i}{2}\mathbf{\Gamma}(\omega)\right)\tilde{\mathbf{v}}(\omega) \nonumber \\
    &+ i\,\mathbf{A}_\epsilon\left[\tilde{\mathbf{v}}(\omega+\omega_m)+\tilde{\mathbf{v}}(\omega-\omega_m)\right]
    + \tilde{\mathbf{F}}_{\mathrm{in}}(\omega)\,,
\end{align}
where $\tilde{\mathbf{v}}(\omega)=\big(\tilde{a}(\omega),\,\tilde{a}^\dagger(\omega)\big)^T$ is the vector of the bosonic operators and $\tilde{\mathbf{F}}_{\mathrm{in}}(\omega) = i / \sqrt{2\hbar\omega_a} \, \tilde{\xi}(\omega) (1 , -1)^T$ denotes the Langevin force vector.

In what follows, adopting the approach of Ref.~\cite{Lamberto2026superradiantquantum}, we explicitly derive the connection between the Langevin forces, the input-output operators, and the bath degrees of freedom.
We begin by taking the continuum limit of $\xi(t)$ in \eqref{eq:xi_ab}, obtaining
\begin{align}
     \xi(t) = \int_0^\infty \sqrt{\frac{\hbar \omega k(\omega)}{2}} \left( \vphantom{e^{-i \omega (t-t_0)}} \right. & \tilde{c}(\omega, t_0) e^{-i \omega (t-t_0)}  \nonumber \\
    & \left. +\, \tilde{c}^\dagger(\omega, t_0) e^{i \omega (t-t_0)} \right) \,d\omega \, .
\end{align}
This expression closely mirrors the time derivative of the input operators defined in \eqref{eq:C_in}, after the unitary transformation of bath bosonic operators $\tilde{c}\to i\,\tilde{c}$, namely
\begin{align}
    \dot{C}_{{\rm in}}(t) = \int_0^\infty \sqrt{\frac{\hbar\omega}{4 \pi}} \left( \vphantom{e^{-i \omega (t-t_0)}} \right. & \tilde{c}(\omega, t_0) e^{-i \omega (t-t_0)}  \nonumber \\
    & \left. +\, \tilde{c}^\dagger(\omega, t_0) e^{i \omega (t-t_0)} \right) \,d\omega \, .
\end{align}
Using the microscopic definition of the damping rate, one finds $k(\omega)=2\gamma(\omega)/\pi$, from which the frequency-domain relation
$\tilde{\xi}(\omega) = -2 i \omega \sqrt{\gamma(\omega)}\,\tilde{C}_{{\rm in}}(\omega)$ follows immediately.

Moreover, we notice that, by explicitly calculating the Fourier transform of $\xi(t)$, one obtains
\begin{align} \label{eq:xi_tilde}
    \tilde{\xi}(\omega) =&\, \frac{1}{\sqrt{2\pi}}\int_{-\infty}^\infty\xi(t) e^{i\omega t}dt =  \nonumber \\
    =& \int_0^{\!\,\infty} \!\!\! \sqrt{\hbar \pi \omega^\prime k(\omega^\prime)}  \left[ \vphantom{\tilde{c}^\dagger_{{\rm in}}}\tilde{c}_{{\rm in}}(\omega^\prime) \delta(\omega - \omega^\prime)\right. \nonumber \\
    &\left. \quad\qquad\qquad\qquad +\, \tilde{c}^\dagger_{{\rm in}}(\omega^\prime) \delta(\omega + \omega^\prime)\right] d\omega^\prime\!\, ,  
\end{align}
where $\tilde{c}_{{\rm in}}(\omega)=\tilde{c}(\omega,t_0)e^{i\omega t_0}$ denotes the bath operator evaluated in the remote past, ideally $t_0\to -\infty$. 

In an analogous manner, we can define the output operators as
$\tilde{c}_{{\rm out}}(\omega)=\tilde{c}(\omega,t_1)e^{i\omega t_1}$, with the bath operators taken in the distant future, $t_1\to +\infty$.
Equation~(\ref{eq:xi_tilde}) reveals that the Dirac delta functions select the annihilation or creation operator of the bath according to the sign of $\omega$. 
In particular, by defining the operator $\check{c}_\mathrm{in}(\omega) = \tilde{c}_\mathrm{in}(\omega) \theta(\omega) + \tilde{c}_\mathrm{in}^\dagger(-\omega) \theta(-\omega)$ as in \cref{eq:c_check} in the main text, we obtain the relation
\begin{equation}
    \tilde{\xi}(\omega)=\sqrt{\hbar \pi \abs{\omega} k(\omega)}\,\check{c}_\mathrm{in}(\omega).
\end{equation}
Finally, the relation between the Langevin forces and the input operators is obtained as
\begin{equation} \label{app:langevin_forces_bosonic}
    \tilde{\bf F}_{\rm in}(\omega) = \frac{i}{\sqrt{2\hbar\omega_a}} \, \tilde{\xi}(\omega) \begin{pmatrix}
        1 \\ -1
    \end{pmatrix} \!= 
    i \sqrt{\frac{\abs{\omega} \gamma_a}{\omega_a}} \, \check{c}_\mathrm{in}(\omega) \begin{pmatrix}
        1 \\ -1
    \end{pmatrix} \,.
\end{equation}
An analogous relation holds for the output fields.


\section{Derivation of the output fields}
\label{sec:derivation_output}

In this Appendix, we explicitly derive the relation between the input and output operators and, consequently, the expressions for the coefficients $k_n (\omega)$ appearing in Eqs.~(\ref{eq:out_in_definition}) and (\ref{eq:out_1_harm}).

To this end, we first rewrite \cref{eq:output_vs_input} in terms of the bosonic input and output operators using \cref{app:langevin_forces_bosonic}. Without loss of generality, we henceforth consider positive output frequencies, $\omega > 0$. This yields
\begin{align} \label{app:out_in_derivation_compact}
    \tilde{c}_{\rm out}(\omega) &= \tilde{c}_{\rm in} (\omega) + i \sqrt{\frac{\omega_a \gamma_a}{\omega}} \, (1, -1) \, \tilde{\mathbf{v}}(\omega) \nonumber \\
    & = \tilde{c}_{\rm in} (\omega) + i \sqrt{\frac{\omega_a \gamma_a}{\omega}} \left[ \tilde{a}(\omega) - \tilde{a}^\dagger (\omega) \right] \, .
\end{align}

The last term in \cref{app:out_in_derivation_compact} contains the contributions induced by the modulation. These can be evaluated by expressing the system operators in terms of the input fields through the inversion of \cref{eq:QLE_input}, namely
\begin{equation} \label{app:system_op_vs_input}
    \tilde{\mathbf{v}} (\omega) = \mathbf{M}^{-1}(\omega) \left\{ \mathbf{A}_\epsilon \left[ \tilde{\mathbf{v}} (\omega \!+\! \omega_m) \!+ \tilde{\mathbf{v}} (\omega \!-\! \omega_m) \right] - i \tilde{\mathbf{F}}_\mathrm{in} (\omega) \right\} .
\end{equation}

This equation can be solved self-consistently by iteratively evaluating the left-hand side at frequencies $(\omega + k \omega_m)$, with $k \in \mathbb{Z}$, and substituting the resulting expressions back into the right-hand side. In practice, this procedure is carried out up to some cutoff $N$, since higher-order contributions become progressively negligible. However, as the system approaches the critical point, these contributions become increasingly comparable to each other, and larger values of $N$ are required to ensure convergence.

The inversion of \cref{app:system_op_vs_input} can be performed semi-analytically by exploiting a useful observation. 
In particular, we first notice that the modulation matrix can be decomposed as $\mathbf{A}_\epsilon = \epsilon (1, -1)^T (1, 1)$, so that the modulation term involves only the scalar quantities $y_{\pm 1}(\omega) = (1,1) \tilde{\mathbf{v}}_{\pm 1} (\omega)$, where (for compactness) we introduce the shorthand notation $O_k (\omega) = O (\omega + k \omega_m)$ for any general quantity $O$. Moreover, the following identities will also prove useful in the upcoming derivation, i.e.,
\begin{subequations} \label{app:identities_M}
\begin{eqnarray}
    (1,-1) \mathbf{M}^{-1}_{k}(\omega) \begin{pmatrix}
        1 \\
        -1
    \end{pmatrix} &=& -\frac{2 (\omega + k\omega_m)}{D_{k}(\omega)} \, ,
    \\
     (1,1) \mathbf{M}^{-1}_{k}(\omega) \begin{pmatrix}
        1 \\
        -1
    \end{pmatrix} &=& -\frac{2 \omega_a}{D_{k}(\omega)}
    \, ,
\end{eqnarray}
\end{subequations}    
where $D(\omega) = \omega^2 + i \gamma_a \omega - \tilde{\omega}_a^2$, as \cref{eq:D_omega} in the main text.
Hence, using the observations above and multiplying \cref{app:system_op_vs_input} on the left by $(1 , 1)$, we obtain the scalar equation
\begin{equation} \label{app:recursive_relation_y}
    y_0(\omega) = -\frac{2 \omega_a}{D_0(\omega)}
    \left\{ \epsilon \left[ \vphantom{\frac{1}{1}} y_{1}(\omega) + y_{-1}(\omega) \right] + y_{\mathrm{in},0}(\omega) \right\} \, ,
\end{equation}
where we have defined
\begin{equation}
    y_{\mathrm{in},k}(\omega) = \sqrt{\frac{\abs{\omega + k \omega_m} \gamma_a}{\omega_a}} \, \check{c}_{\mathrm{in},k}(\omega) \, .
\end{equation}
By evaluating \cref{app:recursive_relation_y} at shifted frequencies $(\omega + k \omega_m)$ up to the $N$-th term, one obtains an equivalent $(2N+1)$-dimensional tridiagonal matrix problem, which can be solved semi-analytically for $y(\omega)$. Once $y_0(\omega)$ has been determined, the quantity $(1,-1) \tilde{\mathbf{v}}_0 (\omega)$ can be obtained by multiplying \cref{app:system_op_vs_input} on the left by $(1,-1)$. This then allows us to express the output operators explicitly in terms of the input ones.

\subsection{$N=1$ case}

As a practical illustration of this procedure, we now derive the analytical expression for the case $N=1$. In this case, \cref{app:recursive_relation_y} leads to the following equivalent tridiagonal system:
\begin{equation} \label{app:tridiagonal_matrix_1harm}
    \setlength{\arraycolsep}{1pt}
    \begin{pmatrix}
        -\frac{D_{1}(\omega)}{2 \omega_a} & -\epsilon & 0 \\
    - \epsilon & -\frac{D_{0}(\omega)}{2 \omega_a} & -\epsilon \\
    0 & -\epsilon & -\frac{D_{-1}(\omega)}{2 \omega_a}
    \end{pmatrix}
    \renewcommand{\arraystretch}{1.2}
    \!\!
    \begin{pmatrix}
        y_{1}(\omega) \\
        y_{0}(\omega) \\
        y_{-1}(\omega)
    \end{pmatrix} \!=\!
    \begin{pmatrix}
        y_{\mathrm{in},1}(\omega) \\
        y_{\mathrm{in},0}(\omega) \\
        y_{\mathrm{in},-1}(\omega)
    \end{pmatrix} \, .
\end{equation}
This system can be solved by inverting the matrix on the left-hand side and subsequently substituting the result into \cref{app:system_op_vs_input}, thereby obtaining \cref{eq:out_1_harm}.

In the case of $N=1$, an alternative (and arguably simpler) approach for obtaining the analytical results is possible.
From \cref{app:recursive_relation_y}, we can compute both the contributions
\begin{equation}
    y_{\pm 1} = -\frac{2 \omega_a}{D_{\pm 1}(\omega)}
    \left\{ \epsilon y_{0}(\omega) + y_{\mathrm{in},\pm 1}(\omega) \right\} \, .
\end{equation}
Substituting these expressions into \cref{app:system_op_vs_input} and collecting the terms, we obtain
\begin{widetext}
\begin{equation}
    \left\{ \mathbf{I} + \left(\! \frac{2\epsilon^2 \omega_a}{D_{1}(\omega)} \!+\! \frac{2\epsilon^2 \omega_a}{D_{-1}(\omega)} \!\right) \mathbf{M}^{-1}_{0}\!(\omega) \! \begin{pmatrix}
        1 \\
        -1
    \end{pmatrix} \! (1,1)\! \right\} \tilde{\mathbf{v}}_{0} (\omega) \! = \mathbf{M}^{-1}_{0}\!(\omega) \! \begin{pmatrix}
        1 \\
        -1
    \end{pmatrix} \!\! \left\{y_{\mathrm{in},0}(\omega) \!-\! \frac{2\epsilon \omega_a}{D_{1}(\omega)} y_{\mathrm{in},1}(\omega) \!-\! \frac{2\epsilon \omega_a}{D_{-1}(\omega)} y_{\mathrm{in},-1}(\omega) \right\} \, . 
\end{equation}
The coefficient of $\tilde{\mathbf{v}}_{0} (\omega)$ can be inverted analytically by employing the identity $\left(\mathbf{I} + \mathbf{U}\mathbf{V}\right)^{-1} = \mathbf{I} - \mathbf{U} (\mathbf{I}+\mathbf{V}\mathbf{U})^{-1} \mathbf{V}$, where $\mathbf{U}$ and $\mathbf{V}$ are two general compatible matrices. This yields
\begin{equation}
    \left\{ \mathbf{I} \!+\! \left(\! \frac{2\epsilon^2 \omega_a}{D_{1}(\omega)} \!+\! \frac{2\epsilon^2 \omega_a}{D_{-1}(\omega)} \!\right) \! \mathbf{M}^{-1}_{0}\!(\omega) \! \begin{pmatrix}
        1 \\
        -1
    \end{pmatrix} \! (1,1)\! \right\}^{\!-1} \!\!= \mathbf{I} - z(\omega)
    \mathbf{M}^{-1}_{0}\!(\omega) \! \begin{pmatrix}
        1 \\
        -1
    \end{pmatrix} (1,1) \, ,
\end{equation}
with the definition $z(\omega) = 2 \epsilon^2 \omega_a D_{0}(\omega) \left[D_{1}(\omega) \!+\! D_{-1}(\omega)\right] / \left\{D_{0}(\omega) D_{1}(\omega) D_{-1}(\omega) - 4 \epsilon^2 \omega_a^2 \left[D_{1}(\omega) \!+\! D_{-1}(\omega)\right] \right\}$.
We can now compute the quantity $(1,-1) \tilde{\mathbf{v}}_{0} (\omega)$, which results in
\begin{eqnarray}
    (1,-1) \tilde{\mathbf{v}}_{0} (\omega) \!&=& (1,-1) \! \left\{ \mathbf{I} - z(\omega)
    \mathbf{M}^{-1}_{0}\!(\omega) \! \begin{pmatrix}
        1 \\
        -1
    \end{pmatrix} (1,1) \right\} \mathbf{M}^{-1}_{0}\!(\omega) \! \begin{pmatrix}
        1 \\
        -1
    \end{pmatrix} \left\{y_{\mathrm{in},0}(\omega) \!-\! \frac{2\epsilon \omega_a}{D_{1}(\omega)} y_{\mathrm{in},1}(\omega) \!-\! \frac{2\epsilon \omega_a}{D_{-1}(\omega)} y_{\mathrm{in},-1}(\omega) \right\} \nonumber \\
    &=& (1,-1) \mathbf{M}^{-1}_{k}(\omega)\! \begin{pmatrix}
        1 \\
        -1
    \end{pmatrix}\!\! \left[ 1 - z(\omega) (1,1) \mathbf{M}^{-1}_{k}(\omega) \! \begin{pmatrix}
        1 \\
        -1
    \end{pmatrix} \right] \! \left\{y_{\mathrm{in},0}(\omega) \!-\! \frac{2\epsilon \omega_a}{D_{1}(\omega)} y_{\mathrm{in},1}(\omega) \!-\! \frac{2\epsilon \omega_a}{D_{-1}(\omega)} y_{\mathrm{in},-1}(\omega) \right\} \nonumber \\
    &=& -\frac{2 \omega}{D_{0}(\omega)} \left[ 1 + \frac{2 \omega_a z(\omega)}{D_{0}(\omega)}\right] \left\{y_{\mathrm{in},0}(\omega) \!-\! \frac{2\epsilon \omega_a}{D_{1}(\omega)} y_{\mathrm{in},1}(\omega) \!-\! \frac{2\epsilon \omega_a}{D_{-1}(\omega)} y_{\mathrm{in},-1}(\omega) \right\} \, ,
\end{eqnarray}
where we used the properties in Eqs.~(\ref{app:identities_M}). Finally, substituting this expression into \cref{app:system_op_vs_input}, we obtain \cref{eq:out_1_harm} of the main paper.
\end{widetext}

\section{Instability boundaries}
\label{sec:instability}

\begin{figure}[t]
    \centering
    \includegraphics[width=\linewidth]{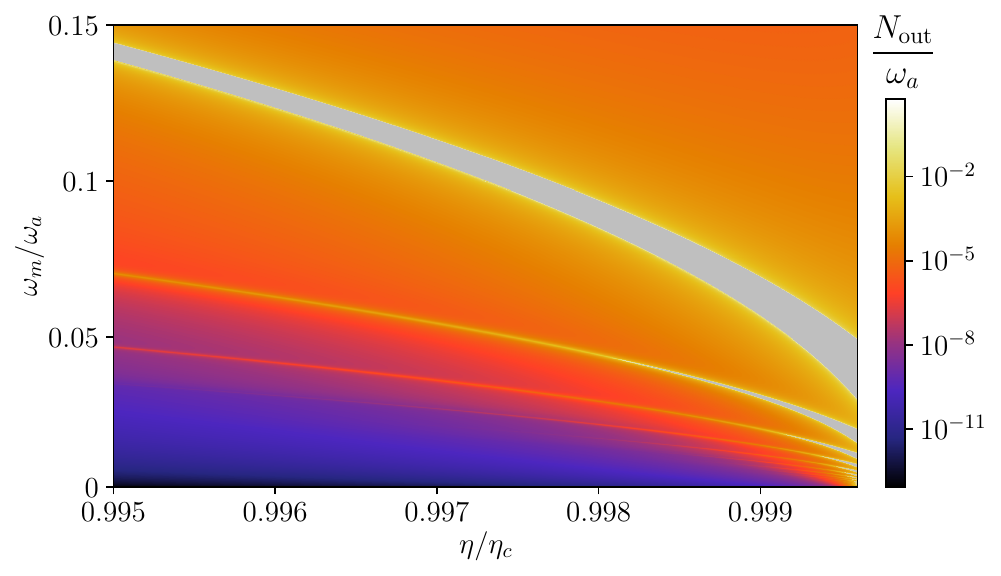}
    \caption{\textbf{Photon flux and associated stability regions.}
    Emitted photon flux as a function of the modulation frequency $\omega_m / \omega_a$ and the normalized coupling $\eta/\eta_c$ near the critical point. Upon approaching criticality, many higher harmonics contribute appreciably to both the emitted photon flux and stability regions which broaden accordingly (shaded grey area), due to the high ratio $\gamma_a / \epsilon$. The parameters used here are: $\gamma_a / \omega_a = 5 \times 10^{-4}$, $\epsilon / \gamma_a = 0.2$.}
    \label{fig:stability_exaggerated}
\end{figure}

The instability regions correspond to those regions in the parameters' space in which the time-domain solutions transition from stable to unstable behavior, namely where the mean values of the position and momentum operators diverge. To determine the instability boundaries in the $\omega_m$--$\epsilon$ plane (as shown, for example, in \cref{fig:photon_flux_spectrum} and \cref{fig:stability_exaggerated}), it proves sufficient to adopt a mean-field approach. 

To this end, in order to work with real-valued equations, it is convenient to express the QLEs for $a(t)$ and $a^\dagger (t)$ in \cref{eq:QLE_time} and its complex conjugate, respectively, in terms of position and momentum operators, namely
\begin{equation} 
    \begin{cases}
    \begin{aligned}
        \dot{x}(t) & = \omega_a p(t) \\
        \rule{0pt}{1ex}
        \dot{p}(t) & = -\omega_a x(t) + 4 \eta(t) x - \gamma p(t) + \sqrt{\frac{2}{\hbar\omega_a}} \xi(t)
    \end{aligned} 
    \end{cases}
\end{equation}
We now take the expectation value of both equations (noting that the noise term has zero mean) and combine them into a single second-order differential equation for the expectation value of the position operator, i.e.,
\begin{equation} \label{app:ODE_position_stability}
    \ddot{\expec{x}} + \gamma \dot{\expec{x}} + \left[\omega_a^2 - 4\omega_a \eta(t)\right] \expec{x} = 0 \, .
\end{equation}
This equation has the form of a real-valued damped Mathieu equation, whose stability can be analyzed by exploiting the fact that, at the instability boundaries, the time-domain solutions become periodic and can therefore be expressed as a Fourier series of the form \cite{Bender1999book, Grimshaw2017nonlinear}
\begin{equation} \label{app:x_expansion_stability}
    x(t) = \frac{a_0}{2} + \sum_{n=1}^{+\infty} \left[ a_n \cos\left(\frac{n \omega_m}{2}t\right) + b_n \sin\left(\frac{n \omega_m}{2}t\right) \right] \, .
\end{equation}
By substituting this definition into \cref{app:ODE_position_stability}, simplifying the resulting trigonometric expressions, and collecting like terms (a procedure known as harmonic balance), we obtain

\begin{widetext}

\begin{align}
    \frac{1}{2} \left(\tilde{\omega}_a^2 a_0 - 4 \epsilon \omega_a a_2\right) + \sum_{n=1}^{+\infty} & \left\{ \vphantom{\frac{1}{1}} \left[ \left(\tilde{\omega}_a^2 -\omega_n^2\right) a_n + \gamma\omega_n b_n - 2 \epsilon \omega_a \left( a_{n-2} + a_{n+2} \right) \right] \cos \left( \omega_n t \right) \right. \nonumber \\
    & \left. \vphantom{\frac{1}{1}} \; + \left[ \left(\tilde{\omega}_a^2 -\omega_n^2\right) b_n - \gamma\omega_n a_n - 2 \epsilon \omega_a \left( b_{n-2} + b_{n+2} \right) \right] \sin \left( \omega_n t \right) \right\} = 0 \, ,
\end{align}
where, for notational convenience, we have defined $\omega_n = n \omega_m / 2$.
Requiring that this equation must hold for all times $t$ implies that each coefficient in the Fourier expansion must vanish independently, leading to the following sets of relations:
\begin{eqnarray}
    & n \;\, \mathrm{even} \;\, (n = 2 k) & \implies
    \left\{
    \begin{aligned}
        & \; \tilde{\omega}_a^2 a_0 - 4 \epsilon \omega_a a_2 = 0 \\
        & \left(\tilde{\omega}_a^2 - \omega_{2k}^2 \right) a_{2k} + \gamma_a \omega_{2k} b_{2k} - 2 \epsilon \omega_a (a_{2k-2} + a_{2k+2}) = 0
              \smash{\raisebox{-0.5\baselineskip}{$\qquad \forall k \ge 1$}}\\
        & \left(\tilde{\omega}_a^2 - \omega_{2k}^2 \right) b_{2k} - \gamma_a \omega_{2k} a_{2k} - 2 \epsilon \omega_a (b_{2k-2} + b_{2k+2}) = 0
    \end{aligned}
    \right.
    \\
    \rule{0pt}{9ex}
    & n \;\; \mathrm{odd} \;\, (n = 2 k + 1) & \implies 
    \left\{
    \begin{aligned}
        & \; \tilde{\omega}_a^2 a_1 + \gamma \omega_1 b_1 - 2 \epsilon \omega_a (a_1 + a_3) = 0 \\
        & \; \tilde{\omega}_a^2 b_1 - \gamma \omega_1 a_1 - 2 \epsilon \omega_a (-b_1 + b_3) = 0 \\
        & \left(\tilde{\omega}_a^2 - \omega_{2k+1}^2 \right) a_{2k+1} + \gamma_a \omega_{2k+1} b_{2k+1} - 2 \epsilon \omega_a (a_{2k-1} + a_{2k+3}) = 0
              \smash{\raisebox{-0.5\baselineskip}{$\qquad \forall k \ge 1$}}\\
        & \left(\tilde{\omega}_a^2 - \omega_{2k+1}^2 \right) b_{2k+1} - \gamma_a \omega_{2k+1} a_{2k+1} - 2 \epsilon \omega_a (b_{2k-1} + b_{2k+3}) = 0
    \end{aligned}
    \right.
\end{eqnarray}
where, without loss of generality, we have assumed $b_0 = 0$, $a_{-1} = a_1$ and $b_{-1} = - b_1$. 
The existence of nontrivial solutions to these equations requires that the system parameters satisfy the condition that the infinite determinants associated with the even and odd coefficients must vanish, which in turn determines the instability boundaries.
As already discussed in \cref{app:derivation_langevin_input} and in the main text, in practice these determinants are truncated at a sufficiently large cutoff $N$.

Figure~\ref{fig:stability_exaggerated} shows the output photon flux together with the corresponding instability regions. Owing to the relatively large ratio $\epsilon / \gamma_a = 0.2$, a higher cutoff $N$ for the number of harmonics is required to accurately compute the photon flux (and consequently the instability boundaries). Accordingly, a larger number of harmonics is visible in \cref{fig:stability_exaggerated} compared to \cref{fig:photon_flux_spectrum}.

\section{Derivation of the Emission and Squeezing Spectra} \label{app:derivation_emission_squeezing_spectra}

We now provide a formal derivation of the analytic expression for the photon flux density presented in the main text:
\begin{eqnarray}
\label{eq:derivation_emission_spec}
n_{\rm out}(\omega)
&=& \frac{1}{2\pi}\, \lim_{T\rightarrow\infty}\frac{1}{T}\int_0^T \! dt \int_0^T \! dt^\prime \,
\big\langle c^\dagger_{\rm out}(t)\, c_{\rm out}(t^\prime) \big\rangle \, e^{-i \omega(t-t^\prime)}
\nonumber \\
&=& \frac{1}{2\pi}\, \lim_{T\rightarrow\infty}\frac{1}{T}
\left[
\int_0^T \! dt \int_0^T \! dt^\prime
\int_0^\infty \! d\omega^\prime \, e^{i(\omega^\prime-\omega)t}
\int_0^\infty \! d\omega^{\prime\prime} \, e^{i(\omega-\omega^{\prime\prime})t^\prime}
\big\langle c^\dagger_{\rm out}(\omega^\prime)\, c_{\rm out}(\omega^{\prime\prime}) \big\rangle
\right]
\nonumber \\
&=& \frac{1}{2}\, \lim_{T \rightarrow \infty}
\left[
\int_0^\infty \!\! d\omega^\prime \int_0^\infty \!\! d\omega^{\prime\prime}\,
e^{i(\omega^{\prime}-\omega^{\prime\prime})T/2}\,
\frac{\sin{\!\left((\omega^\prime-\omega)T/2\right)}}{(\omega^\prime-\omega)T/2}\,
\frac{\sin{\!\left((\omega-\omega^{\prime\prime})T/2\right)}}{(\omega-\omega^{\prime\prime})\pi/2}\,
\big\langle c^\dagger_{\rm out}(\omega^\prime)\, c_{\rm out}(\omega^{\prime\prime}) \big\rangle
\right]
\nonumber \\
&=&\frac{1}{2}\,\lim_{T \rightarrow \infty} \, 
\left[
\int_0^\infty \!\! d\omega^\prime \int_0^\infty \!\! d\omega^{\prime\prime}\, \mathcal{F}_T(\omega^\prime-\omega)\, \delta_T(\omega-\omega^{\prime\prime})\, \big\langle c^\dagger_{\rm out}(\omega^\prime)\, c_{\rm out}(\omega^{\prime\prime}) \big\rangle
\right]
\nonumber \\
&=& \int_0^\infty \! d\omega^\prime \,
\big\langle c^\dagger_{\rm out}(\omega^\prime)\, c_{\rm out}(\omega) \big\rangle \,
\mathcal{F}_{\infty}(\omega^\prime-\omega)\, ,
\end{eqnarray}
where in the final step we relied on a standard limiting representation of the Dirac delta distribution in terms of a sequence of functions
\begin{equation}
\label{eq:dirac_representation}\lim_{T\rightarrow\infty}\delta_T(x)=
\lim_{T\rightarrow \infty} \frac{\sin(x T)}{\pi x} = \delta(x)\, .
\end{equation}
Moreover, we introduced the function 
\begin{equation}
\label{eq:filter_function}
\mathcal{F}_\infty(\alpha)=\lim_{T\rightarrow\infty}\mathcal{F}_T(\alpha)=
\lim_{T\rightarrow \infty}  \frac{\sin(\alpha T)}{\alpha T} e^{i\alpha T/2} =
\begin{cases}
0 & \text{if } \alpha \neq 0\\
1 & \text{if } \alpha = 0
\end{cases} \quad,
\end{equation}
which acts as an effective \textit{frequency selection filter}, in close analogy with a Kronecker delta: in the infinite-time limit it suppresses all contributions with $\alpha\neq 0$ and retains only the resonant terms ($\alpha=0$). Within the theoretical framework developed here, the expectation value 
\(
\big\langle c^\dagger_{\rm out}(\omega^\prime)\, c_{\rm out}(\omega) \big\rangle
\)
is generally nonvanishing even when $\omega \neq \omega^\prime$. Nevertheless, such contributions are effectively suppressed by the factor $\mathcal{F}_\infty(\omega)$. This point is particularly important: without the procedure introduced in \eqref{eq:derivation_emission_spec}, which generalizes the definition of the noise power spectrum to non-strictly stationary conditions, the factor $\mathcal{F}_\infty(\omega)$ would be missing, leading to unphysical contributions (for $\omega \neq \omega^\prime$).

Before deriving the squeezing spectrum, following a procedure analogous to that employed for the emission spectrum, we briefly recall that the quadratures of a continuous multimode field are accessed experimentally via balanced homodyne detection. Assuming the LO to be in a large-amplitude coherent state with well-defined frequency \(\Omega\) and phase \(\theta\), the subsequent intensity measurement yields the output-field quadrature, as defined in the main text
\begin{equation}
    X^{\theta}_{\rm out}(t)=c_{\rm out}(t)\, e^{i(\theta + \Omega t)} + c^\dagger_{\rm out}(t)\, e^{-i(\theta + \Omega t)} \, .
\end{equation}
The corresponding noise power spectrum defines the squeezing spectrum in the frame rotating at \(\Omega\)
\begin{equation}
\label{eq:derivation_ss}
\begin{aligned}
S^\theta_X(\Delta\omega)
&= 1 + \frac{1}{2\pi}\, \lim_{T\rightarrow\infty}\frac{1}{T}\int_0^T \! dt \int_0^T \! dt^\prime \,
\big\langle : \! X^{\theta}_{\rm out}(t)\, X^{\theta}_{\rm out}(t^\prime) \! : \big\rangle \,
e^{-i \Delta\omega(t-t^\prime)}
\\
&= 1
+ \!\!\!\int_0^\infty \! \!\!\!\!d\omega^\prime \, e^{2i\theta}\,
\!\big\langle c_{\rm out}(\Omega-\Delta\omega)\, c_{\rm out}(\omega^\prime) \big\rangle \,\mathcal{F}_{\infty}(\Omega+\Delta\omega-\omega^\prime)\!
+\!\! \int_0^\infty \!\!\!\!\! d\omega^\prime \,
\big\langle c^\dagger_{\rm out}(\omega^\prime)\, c_{\rm out}(\Omega-\Delta\omega) \big\rangle \, \mathcal{F}_{\infty}(\Omega-\Delta\omega-\omega^\prime) 
\\
&\quad +\!\!\!  \int_0^\infty \! \!\!\!\!d\omega^\prime \,
\big\langle c^\dagger_{\rm out}(\Omega+\Delta\omega)\, c_{\rm out}(\omega^\prime) \big\rangle \, \mathcal{F}_{\infty}(\Omega+\Delta\omega-\omega^\prime)\!
+\!\! \int_0^\infty \! \!\!\!\!d\omega^\prime \, e^{-2i\theta}\,
\big\langle c^\dagger_{\rm out}(\omega^\prime)\, c^\dagger_{\rm out}(\Omega-\Delta\omega) \big\rangle \, \mathcal{F}_{\infty}(\Omega+\Delta\omega-\omega^\prime).
\end{aligned}
\end{equation}
Here the frequency \(\Delta\omega\) measured after mixing with the LO is related to the signal-field frequency by \(\omega=\Omega+\Delta\omega\), with the final step again making use of the relations in \eqref{eq:dirac_representation} and \eqref{eq:filter_function}.

As an illustrative example, we consider the case in which 
$\Omega = \omega_m / 2$ is chosen and the analysis is restricted 
to a single harmonic; under these assumptions \eqref{eq:derivation_ss} reduces to
\begin{eqnarray}
      S^\theta_X(\Delta\omega) = 1 &+& \abs{k_{-1}\left(\frac{\omega_m}{2}-\Delta\omega\right)}^2 + \abs{k_{-1}\left(\frac{\omega_m}{2}+\Delta\omega \right)}^2 + 2\operatorname{Re}\left(e^{2i\theta}k_0\left(\frac{\omega_m}{2}-\Delta\omega\right)k_{-1}\left(\frac{\omega_m}{2}+\Delta\omega \right)\right)   , 
\end{eqnarray}
where \(k_0(\omega)\) and \(k_{-1}(\omega)\) are defined in \eqref{eq:out_1_harm} in the main text.

\section{Derivation of the Covariance Matrix}
\label{app:covariance_matrix}

This appendix expands on the procedure introduced in \cref{sec:entanglement}  for constructing the covariance matrix, which is essential for evaluating the 
logarithmic negativity~$\mathcal{N}$. The logarithmic negativity takes positive values for entangled states and can be computed directly from the covariance matrix, defined as
\begin{equation}
    V_{\alpha\beta}
    = \frac{1}{2}\langle r_\alpha r_\beta + r_\beta r_\alpha \rangle \, ,
\end{equation}
where $\mathbf{r} = (q_-,p_-,q_+,p_+)^T$ collects the quadrature operators
\begin{equation}
    q_\pm = \frac{1}{\sqrt{2}}\!\left(b_\pm  
    + b_\pm^\dagger \right) , \qquad
    p_\pm = \frac{i}{\sqrt{2}}\!\left(b_\pm  
    - b_\pm^\dagger \right) ,
\end{equation}
with $b_+ = c_{\rm out}(\omega)$ and $b_- = c_{\rm out}(2\Omega - \omega)$.
We explicitly derive the elements of the covariance matrix $V$
\begin{equation}
V= \frac{1}{2}
\begin{pmatrix}
\langle b_- b_-^\dagger + b_-^\dagger b_- \rangle
& 0
&   \operatorname{Re}\langle b_- b_+ + b_+ b_- \rangle
& - \operatorname{Im}\langle b_- b_+ + b_+ b_-\rangle
\\[4pt]
0
& \langle b_- b_-^\dagger + b_-^\dagger b_- \rangle
& - \operatorname{Im}\langle b_- b_+ + b_+ b_-\rangle
& - \operatorname{Re}\langle b_- b_+ + b_+ b_-\rangle
\\[3pt]
 \operatorname{Re}\langle b_- b_+ + b_+ b_-\rangle
& - \operatorname{Im}\langle b_- b_+ + b_+ b_-\rangle
& \langle b_+ b_+^\dagger + b_+^\dagger b_+ \rangle
& 0
\\[4pt]
-\operatorname{Im}\langle b_- b_+ + b_+ b_-\rangle
& - \operatorname{Re}\langle b_- b_+ + b_+ b_-\rangle
& 0
& \langle b_+ b_+^\dagger + b_+^\dagger b_+ \rangle
\end{pmatrix} \, .
\end{equation}
A direct inspection of this matrix reveals a two-mode squeezing defined by the operator $S(\xi) = \exp\left(\xi^* b_- b_+ - \xi \, b_-^\dagger b_+^\dagger\right)$, with $\xi = r \exp (2i \theta_\mathrm{opt})$. In particular, we can identify $2 \theta_\mathrm{opt} = \arctan(-\operatorname{Im}\langle b_- b_+ + b_+ b_-\rangle / \operatorname{Re}\langle b_- b_+ + b_+ b_-\rangle)$ \cite{Weedbrook2012gaussianRMP}.

As an example, we explicitly consider the first diagonal entry, which is given by
\begin{align}
    V_{11}
    &= \frac{1}{2}\langle b_- b_-^\dagger + b_-^\dagger b_- \rangle \nonumber \\
    &= \frac{1}{2} \Big(
        |k_{-1}(\omega_m - \omega)|^2\!\left[1 + 2\bar{n}_{\mathrm{in}}(\omega)\right] + |k_{0}(\omega_m - \omega)|^2\!\left[1 + 2\bar{n}_{\mathrm{in}}(\omega_m - \omega)\right]
        \theta(\omega_m - \omega) \nonumber \\
    &\qquad\; + |k_{1}(\omega_m - \omega)|^2\!\left[1 + 2\bar{n}_{\mathrm{in}}(2\omega_m - \omega)\right]
        \theta(2\omega_m - \omega)
        \Big) ,
\end{align}
where, for simplicity, we truncated the expansion of the output annihilation operator 
in \cref{eq:out_in_definition} to the first harmonic ($N=1$), and included thermal 
contributions.

Once the covariance matrix for the selected pair of modes is constructed, the 
logarithmic negativity follows as
\begin{equation}
    \mathcal{N} = \max\!\left[0,\; -\log\!\left(2\nu_-\right)\right] ,
\end{equation}
where the definition of $\nu_-$ is provided in the main text.

\end{widetext}

\bibliography{bibliography}

\end{document}